\documentclass[%
amssymb, amsmath,
aps,cha,
superscriptaddress
]{revtex4-2}
\usepackage[colorlinks=true,linkcolor=blue]{hyperref}%
\usepackage{physics}
\usepackage{graphicx}
\usepackage{lipsum}
\usepackage{mwe}
\usepackage{caption}
\usepackage{amssymb}
\usepackage{color}
\usepackage{caption}
\usepackage{subcaption}
\usepackage{tikz}
\usepackage[capitalize]{cleveref}
\usetikzlibrary{quantikz2}
\usepackage{multirow}

\begin{document}

\title{Shell-model study of $^{58}$Ni using quantum computing algorithm}%

\author{Bharti Bhoy}
\email{b.bhoy@surrey.ac.uk}
\affiliation{Department of Physics, University of Surrey, Guildford, Surrey, GU2 7XH, United Kingdom}

\author{Paul Stevenson}
\email{p.stevenson@surrey.ac.uk}
\affiliation{Department of Physics, University of Surrey, Guildford, Surrey, GU2 7XH, United Kingdom}

\begin{abstract}

This study presents a simulated quantum computing approach for the investigation into the shell-model energy levels of $^{58}$Ni through the application of the variational eigensolver (VQE) method in combination with a problem-specific ansatz. The primary objective is to achieve a fully accurate low-lying energy spectrum of $^{58}$Ni. The chosen isotope, $^{58}$Ni is particularly interesting in nuclear physics through its role in astrophysical reactions while also being a simple but not-trivial nucleus for shell-model study, it being two particles outside a closed shell. Our ansatz, along with the VQE method are shown to be able to reproduce exact energy values for the ground state and first and second excited states. We compare a classical shell model code, the values obtained by diagonalization of the Hamiltonian after qubit mapping, and a noiseless simulated ansatz+VQE simulation. The exact agreement between classical and qubit-mapped diagonalization shows the correctness of our method, and the high accuracy of the simulation means that the ansatz is suitable to allow a full reconstruction of the full nuclear wave function.


\end{abstract}


\maketitle

\section{Introduction}
Atomic nuclei are systems of interacting spin-1/2 fermions. As such, their simulation is a strong candidate for a possibly revolutionary treatment using quantum computation, as has been explored in nuclei \cite{visnak_quantum_2015,visnak_quantum_2017,dumitrescu_cloud_2018,gibbs_exploiting_2024,illa_quantum_2023,robin_quantum_2023,roggero_preparation_2020,stevenson_p_d_comments_2023,zhang_selected_2021,stetcu_variational_2021,siwach_quantum_2021,robbins_benchmarking_2021,chikaoka_quantum_2022,li_solving_2024,li_quantum_2024,cervia_lipkin_2021} and other many-fermion systems \cite{brown_using_2010,RevModPhys.92.015003,ma_quantum_2020}. A standard microscopic (treating nucleons as the degree of freedom) method to interpret observed states of nuclei is the configuration interaction nuclear shell model \cite{mcgrory_large-scale_1980,caurier_shell_2005,suhonen} in which nucleons interact in a defined single-particle basis to produce correlated eigenstates of a model Hamiltonian. Finding the wavefunctions and consequent properties of these eigenstates is the job of a shell model practitioner;  a job which has already been explored on real and simulated quantum computers in the case of the lightest nuclei \cite{6He,6Li} with various aspects of the specific algorithms needed for quantum computation examined \cite{perez-obiol_nuclear_2023,sarma}.

If quantum computation is to form a future tool for large-scale shell model calculations on near-term hardware, much work will be needed to understand appropriate algorithms to find the eigenstates and to deal with larger problem sizes. Here we take a step in the latter direction by exploring a heavier system than previously explored, looking at a nucleus ($^{58}$Ni) with valence particles in the $fp$-shell. In our exploration, a simulated variational quantum eigensolver (VQE) algorithm is used to compare aspects of the method, in comparison with an exact classical shell model calculation.  The choice of $^{58}$Ni is primarily motivated as being the simplest two-valence-particle system in the $fp$ shell, and one with which we can compare our quantum simulation calculations with the classical shell model.  This isotope is also of particular interest in astrophysics in the $s$-process in AGB stars \cite{massimi_n_tof_2022,the_n_tof_collaboration_experimental_2014,guber_astrophysical_2010},  as a well-studied nuclear for use as a nuclear data benchmark \cite{alhassan_bayesian_2024}, and as a nuclear material \cite{luneville_impact_2018}.



The primary objective of our research is to achieve a high level of precision in determining the low-lying energy levels of $^{58}$Ni, by constructing a problem-based ansatz able to reproduce exact results based on the shell-model interaction JUN45 \cite{jun45}. We give a survey of the shell model and quantum computation methods in section \ref{sec:theory}, before presenting results in section \ref{sec:results} and some concluding remarks in section \ref{sec:conclusion}.

\section{Theoretical framework}
\label{sec:theory}

In this study, we give details of our methods to investigate the properties of the $^{58}$Ni nucleus using quantum computing simulation techniques, along with a classical shell model comparison.  We take the model space comprising orbitals $1p_{3/2}$, $0f_{5/2}$, and $1p_{1/2}$ above the $^{56}$Ni core. This section describes the Hamiltonian employed along with its adaptation for quantum computing simulation. We present the tailored ansatz as appropriate to reach the ground state of the system. Furthermore, we discuss the application of the Variational Quantum Eigensolver (VQE) algorithm and associated optimizers for energy minimization. The implementation of excitation operators, crucial for constructing quantum circuits, is detailed.

\subsection{Model space and interactions}

The shell model Hamiltonian can be expressed in terms of second quantization, where creation and annihilation operators act on states representing single-particle wavefunctions. The shell model Hamiltonian $H$ consists of the single-particle energy term and the two-body interaction term:
\begin{equation}
H = \sum_{i=1} \epsilon_i \hat{a}^\dagger_i \hat{a}_i + \frac{1}{2} \sum_{i,j,k,l} V_{ijkl} \hat{a}^\dagger_i \hat{a}^\dagger_j \hat{a}_l \hat{a}_k
\end{equation}

Here, $\hat{a}^\dagger_i $ and $\hat{a}_i $ are the creation and annihilation operators, $\epsilon_i $ is the single-particle energy, $V_{ijkl} $ represents the two-body matrix elements (TBMEs) describing interactions between nucleons in states $|i\rangle $, $|j\rangle $, $|k\rangle $, and $|l\rangle $. The nucleon state $|i\rangle $ is characterized by quantum numbers: radial, orbital angular momentum, total spin, and projections of spin and isospin as $|n, l, j, m_{\alpha}, m_{t\alpha}\rangle $. We use the JUN45 \cite{jun45} interaction, which consists of  the $1p_{3/2}$, $0f_{5/2}$, $1p_{1/2}$ and $0g_{9/2}$ orbital in the usual spectroscopic notation. In this work, we have neglected the $0g_{9/2}$ orbital, choosing only $fp$ model space for a simpler approach comprising 12 single-particle states, which remain appropriate for the lowest-energy states.   Since we consider the case of $^{58}$Ni, consisting of a core plus two valence neutrons, we only consider the neutron single-particle states. We represent these orbitals with 12 qubits, one for each orbital, as detailed in Table \ref{q_mapping}.

Since matrix elements of shell model interactions are generally given in the $J$-scheme in which single-particle states are coupled to good total angular momentum $J$, while a quantum computation implementation relies on a mapping of the uncoupled single-particle states (the ``$M$-scheme") to each qubit, a transformation from the $J$-scheme to the $M$-scheme is applied. The transformation is performed using \cite{suhonen}

\begin{align}
\label{eq:m_scheme}
\overline{v}_{\alpha\beta\gamma\delta} = &\sum_{JM,TM_{T}} \left[ {N_{ab}(JT)N_{cd}(JT)} \right]^{-1}  \left(  j_{a} m_{\alpha}  j_{b}  m_{\beta} | JM \right)   \left(\frac{1}{2} m_{t_{\alpha}} \frac{1}{2} m_{t_{\beta}} \Big| T M_{T} \right) \\
&\times\left(  j_{c} m_{\gamma}  j_{d} m_{\delta} | JM \right) \left(  \frac{1}{2} m_{t_{\gamma}} \frac{1}{2} m_{t_{\delta}} \Big| TM_{T} \right) \langle ab; JT|V|cd; JT \rangle \nonumber
\end{align}

The normalization constants $N_{ab}(JT)$ and $N_{cd}(JT)$ are defined as follows:

\begin{align}
N_{ab}(JT) &= \sqrt{\frac{1 - \delta_{ab}(-1)^{J+T}}{1 + \delta_{ab}}} \\
N_{cd}(JT) &= \sqrt{\frac{1 - \delta_{cd}(-1)^{J+T}}{1 + \delta_{cd}}}
\end{align}

Here, we follow the notation of Suhonen \cite{suhonen}, in which a,b,c,d are the single-particle orbitals $ | a \rangle = |n_{\alpha}, l_{\alpha}, j_{\alpha} \rangle $, $|b\rangle=|n_\beta,l_\beta,j_\beta\rangle$, $|c\rangle=|n_\gamma,l_\gamma,j_\gamma\rangle$, $|d\rangle=|n_\delta,l_\delta,j_\delta\rangle$ carrying no projection ($m$, $m_t$) labels. The Greek letters $ \alpha, \beta, \gamma, \delta $ are the nucleon states as a complete set of quantum numbers containing the isospin parts $ | \alpha \rangle = |n_{\alpha}, l_{\alpha}, j_{\alpha}, m_{\alpha}, m_{t\alpha} \rangle $ etc.  In (\ref{eq:m_scheme}),  the four quantities inside round brackets are the Clebsch–Gordan coefficient of angular momenta and isospins, $ \langle ab; JT|V|cd; JT \rangle $ is the $J$-scheme TBMEs.  
The quantum numbers represented by Greek letter indices are tabulated in Table \ref{q_mapping}, and the $m$-scheme TBMEs ($\overline{v}_{\alpha\beta\gamma\delta}$) resulting from equation (\ref{eq:m_scheme}) are given in Appendix \ref{appendix:A}, as transformed from the $J$-scheme values tabulated in \cite{jun45}.


\begin{table}[h]
\centering
    \caption{Qubit mapping for the valence space of $^{58}\text{Ni}$. Each qubit corresponds to a specific combination of  $j$ the total angular momentum, $m_{\alpha}$ projection on the $z$ axis, and $m_{t\alpha}$ the third component of the isospin, is fixed at 1/2 for all qubits to represent the neutron.}
    \label{q_mapping}
\begin{tabular}{ccccccccccccc} 
\hline
$\mathrm{qubit}(\alpha)$ & ~0~ & ~1~ & ~2~ & ~3~ & ~4~ & ~5~ & ~6~ & ~7~ & ~8~ & ~9~ & ~10~ & ~11 \\ 
\hline \hline
$n$ & 1 & 1 & 1 & 1 & 0 & 0 & 0 & 0 & 0 & 0 & 1 & 1 \\ 
$l$ & 1 & 1 & 1 & 1 & 3 & 3 & 3 & 3 & 3 & 3 & 1 & 1 \\ 
$j$ & 3/2 & 3/2 & 3/2 & 3/2 & 5/2 & 5/2 & 5/2 & 5/2 & 5/2 & 5/2 & 1/2 & 1/2 \\ 
$m_{\alpha}$ & $-3/2$ & $3/2$ & $-1/2$ & $1/2$ & $-5/2$ & $5/2$ & $-3/2$ & $3/2$ & $-1/2$ & $1/2$ & $-1/2$ & $1/2$ \\ 
\hline
\end{tabular}
\end{table}

\subsection{Jordan-Wigner qubit mapping}

The Jordan-Wigner transformation \cite{JW} maps fermionic creation and annihilation operators onto Pauli spin matrices. Since the states of the qubits of quantum computers can be written as spinors, with Pauli spin matrices forming a complete set with which to define Hermitian operators acting on the qubits, this JW mapping is often used to translate general many-fermion Hamiltonians to a qubit representation. We follow this method, while noting that other mapping schemes exist \cite{PhysRevC.104.034301}.

The Jordan-Wigner transformation is defined as follows. Consider a system of $ N $ fermionic modes (e.g., nucleons in different energy levels in a nucleus). Each mode can be occupied (state $ |1\rangle $) or unoccupied (state $ |0\rangle $). The occupation number operators are denoted as $ \hat{n}_j = a^\dagger_j a_j $, where $ a^\dagger_j $ and $ a_j $ are the creation and annihilation operators for fermions in mode $ j $, respectively. The Jordan-Wigner transformation maps these fermionic operators to spin-$\frac{1}{2}$ (qubit) operators:

\begin{equation}
a_j \mapsto \left( \prod_{k=1}^{j-1} \sigma^z_k \right) \sigma^-_j, \quad a^\dagger_j \mapsto \left( \prod_{k=1}^{j-1} \sigma^z_k \right) \sigma^+_j,
\end{equation}

where $ \sigma^+_j = \frac{1}{2}(\sigma^x_j + i \sigma^y_j) $ and $ \sigma^-_j = \frac{1}{2}(\sigma^x_j - i \sigma^y_j) $ are the spin raising and lowering operators for the $ j $-th qubit and $\sigma^x_j$ and $\sigma_j^y$ are the Pauli X and Y operator, respectively, and $ \sigma^z_j $ is the Pauli Z operator for the $ j $-th qubit. The product $ \prod_{k=1}^{j-1} \sigma^z_k $ ensures that the fermionic anti-commutation relations are preserved.

The Jordan-Wigner transformation and similar methods, like the Bravyi-Kitaev transformation \cite{BK}, allow a successful mapping of fermionic Hamiltonians onto qubit degrees of freedom. Each possible transformation has its strengths and weaknesses in terms of resulting Hamiltonian complexity, circuit depth and locality/non-locality of representation.  We choose the JW method as a standard against which future comparisons can be made.

\subsection{ VQE and the wavefunction ansatz }
\label{subsec:ansatz}

The Variational Quantum Eigensolver (VQE) \cite{peruzzo,cerezo2020variationalreview,fedorov_vqe_2022} is a hybrid quantum-classical algorithm proposed as a promising approach for finding ground states in many-body quantum systems.  It has been applied in different guises in nuclear physics problems \cite{hobday_variance_2023,cervia_lipkin_2021,dumitrescu_cloud_2018,robin_quantum_2023,stetcu_variational_2021,robbins_benchmarking_2021} thanks to its ability to be implemented, in some cases, on current quantum computing hardware.  The basis of the VQE algorithm is to have a parameterized wave function ansatz on a quantum computer, in which the parameters are free variables; the expectation value of the Hamiltonian in its transformed Pauli matrix form is evaluated on the quantum computer, then a classical algorithm adjusts the parameters of the wave function ansatz to seek a minimum of the Hamiltonian expectation value.  This is done through a repeated cycle of expectation value measurements on the quantum computer and directed parameter adjustments on the side of the classical algorithm. An appropriate ansatz (a guess for the wave function) must be chosen. In nuclear physics, this could be inspired by traditional nuclear models or techniques, like the Unitary Coupled Cluster (UCC) method \cite{coester_bound_1958,kummel_many-fermion_1978}, or by problem-specific considerations.  

In our study, we choose a problem-inspired ansatz in which Pauli $X$ gates initially create the correct number of particles in a state with quantum numbers appropriate to the desired target state.  For the ground state of $^{58}$Ni, characterized by $J = 0$ (and hence $M = 0$), the initial state takes the form of a product state, such as $|0, 1\rangle \equiv |000000000011\rangle$ - i.e. the states $|n=1,l=1,j=3/2,m=-3/2\rangle$ and $|n=1,l=1,j=3/2,m=3/2\rangle$ are occupied. This state is straightforward to prepare as each qubit can be addressed individually and the default initial $|0\rangle$ state of a qubit can be flipped to $|1\rangle$ by application of an $X$ gate. 

Moving to the first excited state of $^{58}$Ni, with $J = 2$, the $M$ value can take values  $-2$, $-1$, $0$, $1$, or $2$.  Opting for simplicity, we choose the $|1, 3\rangle\equiv|000000001010\rangle$ combining $|n=1,l=1,j=3/2,m=3/2\rangle$ and $|n=1,l=1,j=3/2,m=1/2\rangle$ as the reference state among all the other possible combinations of qubits for $J$ = 2. Similarly, for the second excited state with $J = 4$, allowing $M$ to take values like $-4$, $-3$, $-2$, $-1$, $0$, $1$, $2$, $3$, or $4$, we choose the extremal value to avoid contamination with lower $J$ contributions.  Here, we have only two combination for $ M = 4 $, and we opt for the $|1, 5\rangle\equiv|00000100010\rangle$ as the reference state, corresponding to states $|n=1,l=1,j=3/2,m=3/2\rangle$ and $|n=0,l=3,j=5/2,m=5/2\rangle$.

 Having created a simple single configuration reference state for a given target state, parameterized operators are applied to explore the other configurations expected to contribute to the given $M$-valued wave function.  We use parameterized Givens rotation gates \cite{givens_rot} to access selected configurations in $^{58}$Ni with correct $M$ values.  These gates are particle number preserving, keeping us in the correct Fock space, and are suitably parameterized with enough freedom to find the lowest energy configuration.  The excitations in the quantum simulations have been made using the single and double Givens excitation gates. 
 
\begin{figure}[h]
     \centering
         \includegraphics[width=0.45\textwidth]{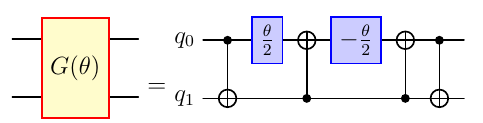}
         \caption{Single-excitation gate in terms of single-qubit Pauli Y rotations ($\pm \frac{\theta}{2}$) and CNOTs \cite{givens_rot}.}
        \label{Single_excitation}
\end{figure}

 Single excitations, shown in Figure \ref{Single_excitation}, involve the application of creation and annihilation operators to excite a single particle from one orbital to another, as represented by $a^\dagger_i a_j$. Since the nucleon carries an intrinsic angular momentum (spin), the single excitation process is applied so that it conserves $m_{\alpha}$ by choosing the individual spins of the nucleons involved in the excitation.

\begin{figure}[h]
     \centering
         \includegraphics[width=\textwidth]{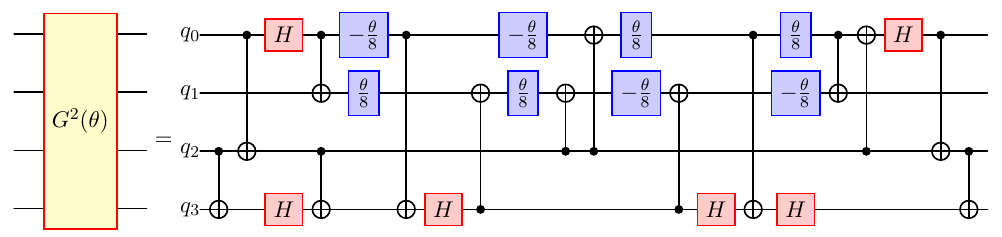}
         \caption{Double-excitation gate in terms of single-qubit Pauli Y rotations ($\pm \frac{\theta}{8}$) and CNOTs \cite{G2,givens_rot}.}
         \label{Double_excitation}
\end{figure}

Double excitations (Figure \ref{Double_excitation}), on the other hand, simultaneously promote two neutrons from their respective orbitals to other orbitals, characterized by expressions like $a^\dagger_i a^\dagger_j a_k a_l$. The set of double excitation processes is chosen so that the total angular momentum projection along the $z$-axis remains constant by accounting for the combined spins of the two nucleons undergoing excitation.

\begin{figure}[h!]
     \centering
     \begin{subfigure}{0.63\textwidth}
         \centering
         \includegraphics[width=\textwidth]{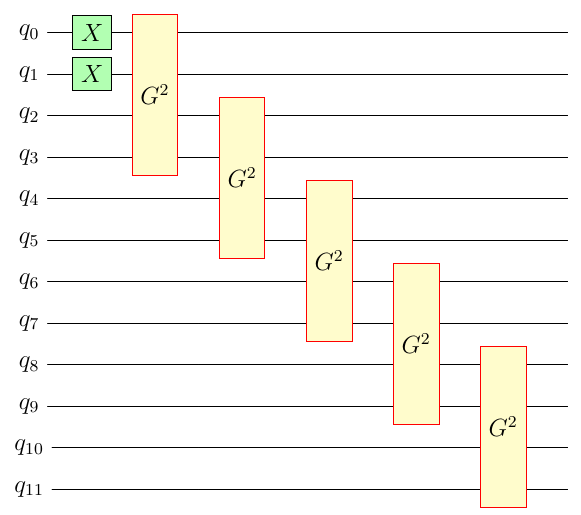}
         \caption{}
         \label{ground_ansatz}
     \end{subfigure}
     \hfill
     \begin{subfigure}{0.30\textwidth}
         \centering
         \includegraphics[width=\textwidth]{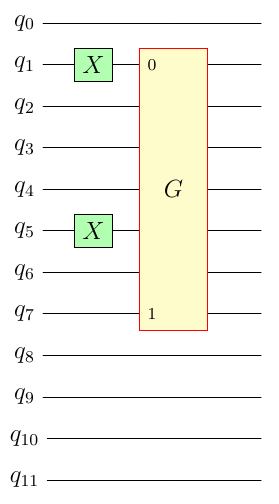}
         \caption{}
         \label{4p_ansatz}
     \end{subfigure}
     \hfill
     \begin{subfigure}{0.85\textwidth}
         \centering
         \includegraphics[width=\textwidth]{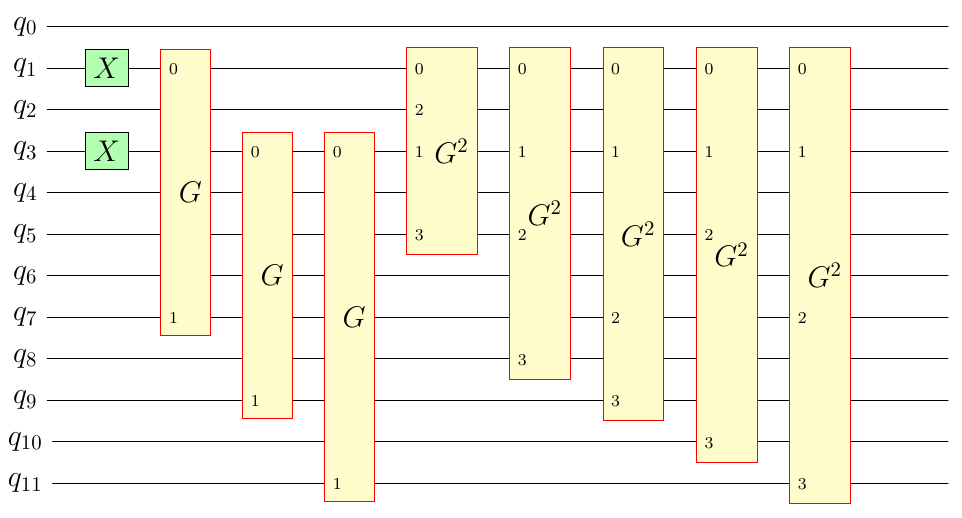}
         \caption{}
         \label{2p_ansatz}
     \end{subfigure}
        \caption{The quantum circuit (a) ground state, (b) second-excited state, and (c) first-excited state used for the VQE simulations for $^{58}$Ni. As mentioned in the main text, the circuits used are composed of a set of initial state preparation gates and single and double-excitation gates.  Numbers on gates indicate qubits between which gates are acting.}
        \label{ansatz}
\end{figure}

The prepared ansatz for the ground state, first excited state, and second excited state is represented in Figure \ref{ansatz}.  Note that while all qubits are active in the ground state wave function, only three are active in the $J=4$ state as only they can give a total projection $M=4$, while nine are active in the $J=2$ state.

The ground state ansatz makes use of the understanding that the ground state wave function will not break pairs, and hence make pairwise excitations from the starting configuration always to $\pm m$ levels into a common $\{nlj\}$ level. The double excitations applied in the circuit are interposed between the qubits $\{0,1,2,3\}$, $\{2,3,4,5\}$, $\{4,5,6,7\}$, $\{6,7,8,9\}$, $\{8,9,10,11\}$, which are the complete set satisfying our non-pair-breaking requirement for $M=0$.

For the first excited state, our approach combines pair-wise excitation and pair-breaking mechanisms, utilizing both single-excitation and double-excitation gates as illustrated in Figure \ref{2p_ansatz}. We initialize the system in the state $|1, 3\rangle$ to explore the first excited state. Specifically, we apply five double excitations: $\{1,3,5,8\}$, $\{1,3,5,10\}$, $\{1,3,2,5\}$, $\{1,3,7,9\}$, $\{1,3,7,11\}$, alongside three single excitations: $\{3,9\}$, $\{3,11\}$, and $\{1,7\}$ among the qubits. 

In the model space for $^{58}$Ni, only a limited number of combinations are feasible for the second excited state ($4^+$). In this context, we have formulated the ansatz employing solely a single excitation, as illustrated in Figure \ref{4p_ansatz}. The system is initialized in the state $|1, 5\rangle$, explicitly applying the single excitation $\{1,7\}$ among the qubits.

\subsection{Classical optimizers}
\label{subsec:optimizer}

Classical optimizers play a key role in hybrid quantum-classical algorithms, such as the Variational Quantum Eigensolver (VQE). The role of the classical optimizer is to adjust the parameters in the wave function ansatz - in our case the angles in each Givens rotation - to optimize the objective function - in our case minimization of the energy expectation value. Here is an overview of the different classical optimizers used in this work, which we take the built-in version from the IBM qiskit package \cite{Qiskit}.

\subsubsection{COBYLA}

The COBYLA (Constrained Optimization BY Linear Approximations) \cite{cobyla} optimizer is a derivative-free method suited for nonlinear optimization problems (which are often present in nuclear physics models), especially where gradients are unavailable or costly to compute. It employs linear approximations to construct a model of the objective function, optimizing within a dynamically adjusted trust region. COBYLA is effective for handling moderate-sized problems and constraints, making it a practical choice in various applications, but may be less efficient for large-scale optimization tasks or when gradient-based methods are applicable.

\subsubsection{SLSQP}

The SLSQP (Sequential Least Squares Programming) \cite{slsqp} optimizer is a gradient-based method effective for solving optimization problems with both linear and nonlinear constraints. It operates by iteratively solving a series of quadratic subproblems, using gradient information to refine solutions at each step. SLSQP is particularly valuable in scenarios where precision in parameter control and cohesion to constraints are crucial, as in nuclear many-body problems. While efficient for problems with available and accurate gradient information, its performance can be limited in cases where such data is difficult to compute or unreliable.

\subsubsection{SPSA}

The SPSA (Simultaneous Perturbation Stochastic Approximation) \cite{spsa} optimizer efficiently tackles high-dimensional optimization by using stochastic methods to approximate gradients, perturbing all parameters simultaneously. This reduces computational demands, especially in problems with many variables. SPSA excels in complex, noisy scenarios, although it might need more iterations and careful tuning to converge. Its suitability for complex nuclear many-body problems comes from its ability to manage numerous variables and work effectively without exact gradient information.

\subsubsection{Gradient Descent}

Gradient Descent is a basic gradient-based optimization method that minimizes functions by iteratively adjusting parameters against the gradient of the objective functions. While effective for problems with well-defined gradients, its performance may suffer with local minima, non-convexities, or noisy gradients. Commonly used in machine learning, its applicability in nuclear many-body problems is limited due to difficulties in obtaining precise gradients and navigating complex energy landscapes. Despite these challenges, it can be effective if accurate gradients are available.

\section{Results}
\label{sec:results}

This section presents the results obtained from  the previously prepared parameterized quantum circuit in Sec. \ref{subsec:ansatz}. We employ the four optimization algorithms given in the previous section: COBYLA, SLSQP, SPSA, and Gradient Descent. We investigate the performance and convergence behaviour of the different optimization algorithms in VQE simulations for computing the expectation value of $^{58}$Ni. The optimization algorithms each navigate the parameter space adeptly, at least in the noiseless simulations we present here, avoiding barren plateaus \cite{PRXQuantum.3.010313} at the level of noiseless simulation. Table \ref{Circ_prop} shows a detailed breakdown of the properties associated with each circuit.  Through the use of our problem-specific ansatz, the number of gates is kept low compared with more general ansatzes \cite{perez-obiol_nuclear_2023}.  We have not imposed nearest-neighbour qubit interaction which is a requirement for some quantum computing hardware, and for the $J=2$ and $J=4$ states, doing this would increase the gate count, though this increase could be minimized with careful re-mapping of single-particle states to qubits. Details of the performance of each optimizer are given in the following section.


\subsection{Ground state}

\begin{figure}[h]
\includegraphics[width=\textwidth,height=\textheight,keepaspectratio]{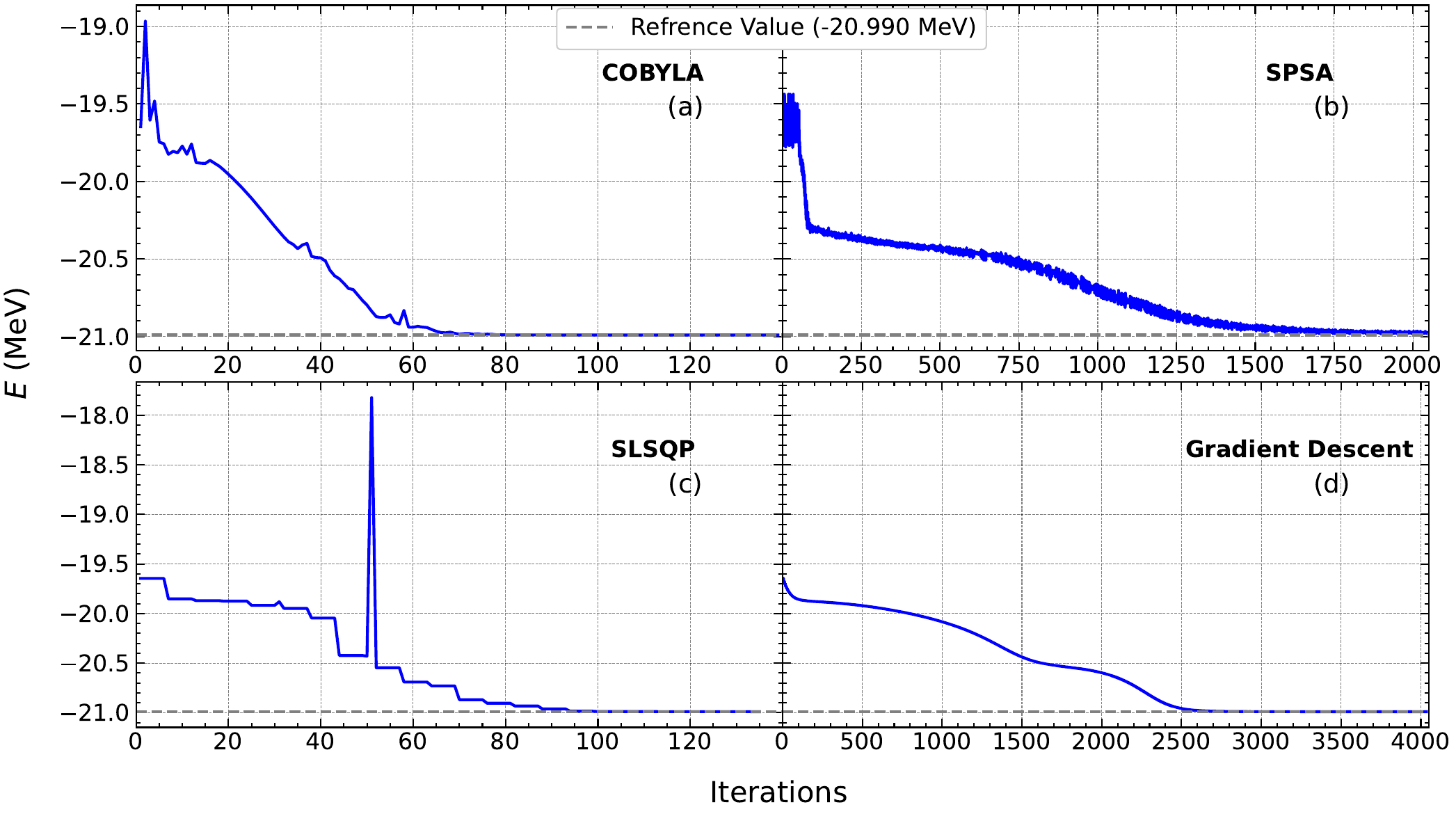}
   \caption{ Different optimizers convergence results for 12 Qubit $0^{+}$ ground state.}
  \label{12_ground}
\end{figure}

\begin{figure}[h!]
    \centering   \includegraphics[width=0.90\textwidth,height=0.43\textheight]{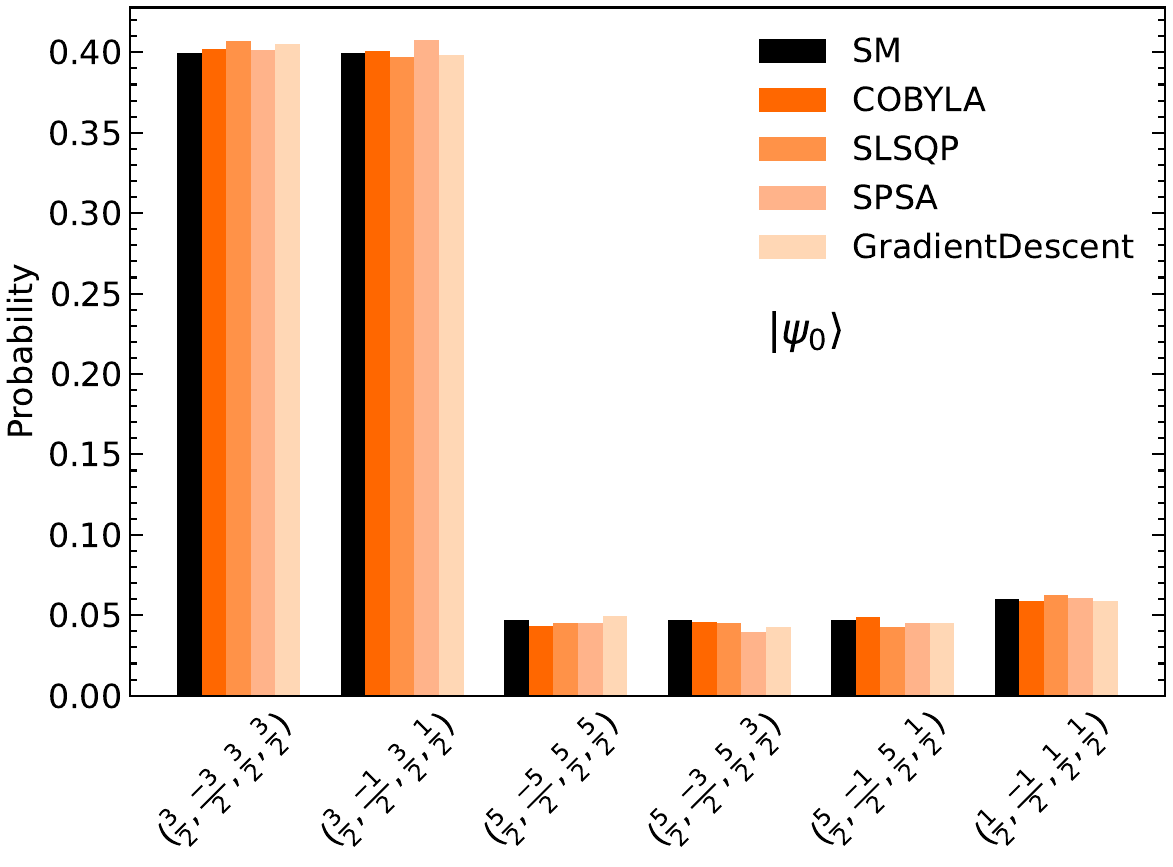}
    \caption{Components of the ground state wave-function with different optimizers in comparison with the Shell model. The $x$-axis labels represent $(j_1,m_{\alpha_1},j_2,m_{\alpha_2})$ and $(m_{\alpha_1}+m_{\alpha_2}) = 0$.}
    \label{G_wf}
\end{figure}

\begin{figure}[h]
\includegraphics[width=\textwidth,height=\textheight,keepaspectratio]{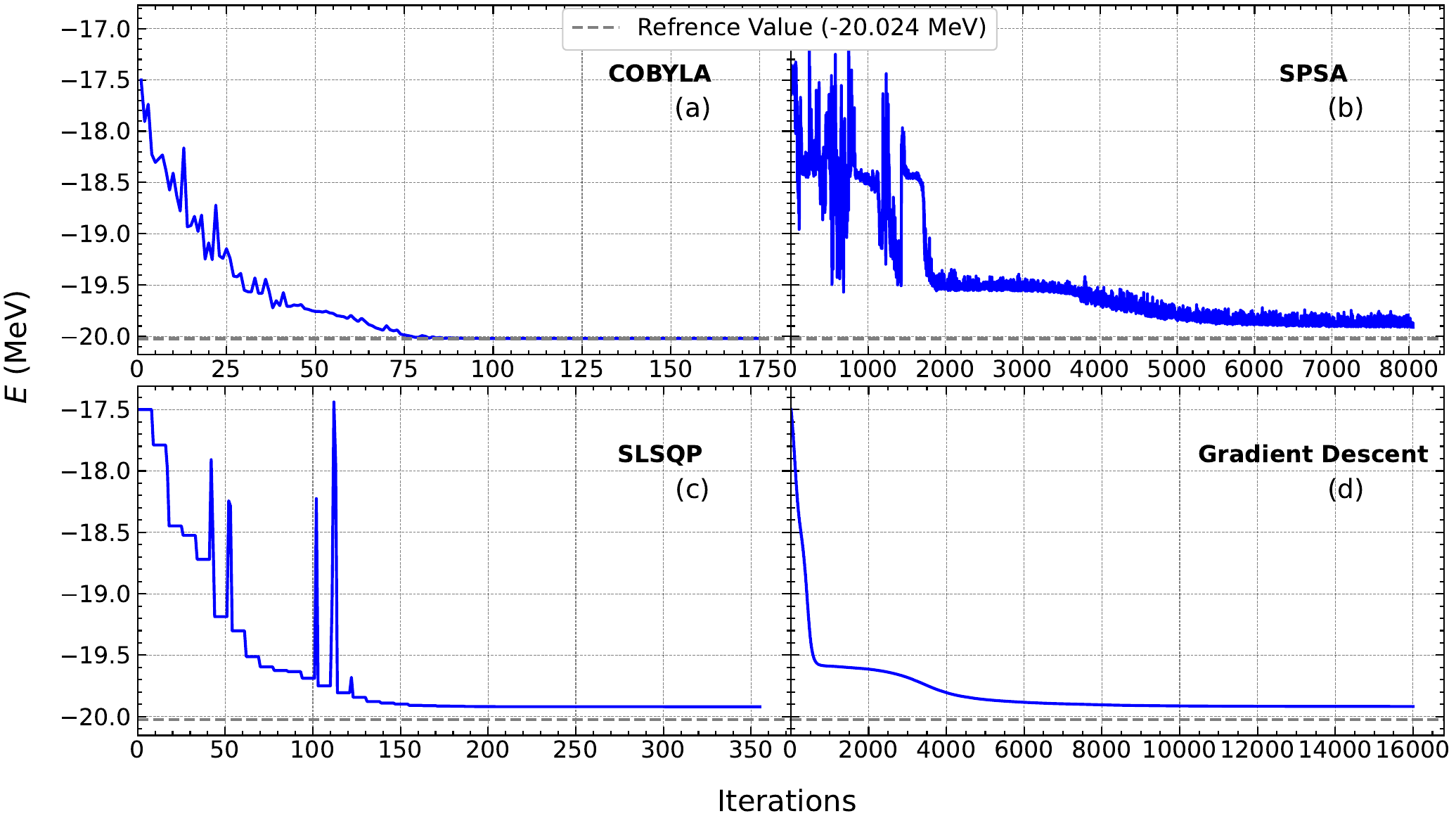}
   \caption{ Different optimizers convergence results for 12 Qubit $2^{+}$ ground state.}
  \label{12_excited_2p}
\end{figure}

\begin{figure}[h!]
    \centering
    \includegraphics[width=0.90\textwidth,height=0.43\textheight]{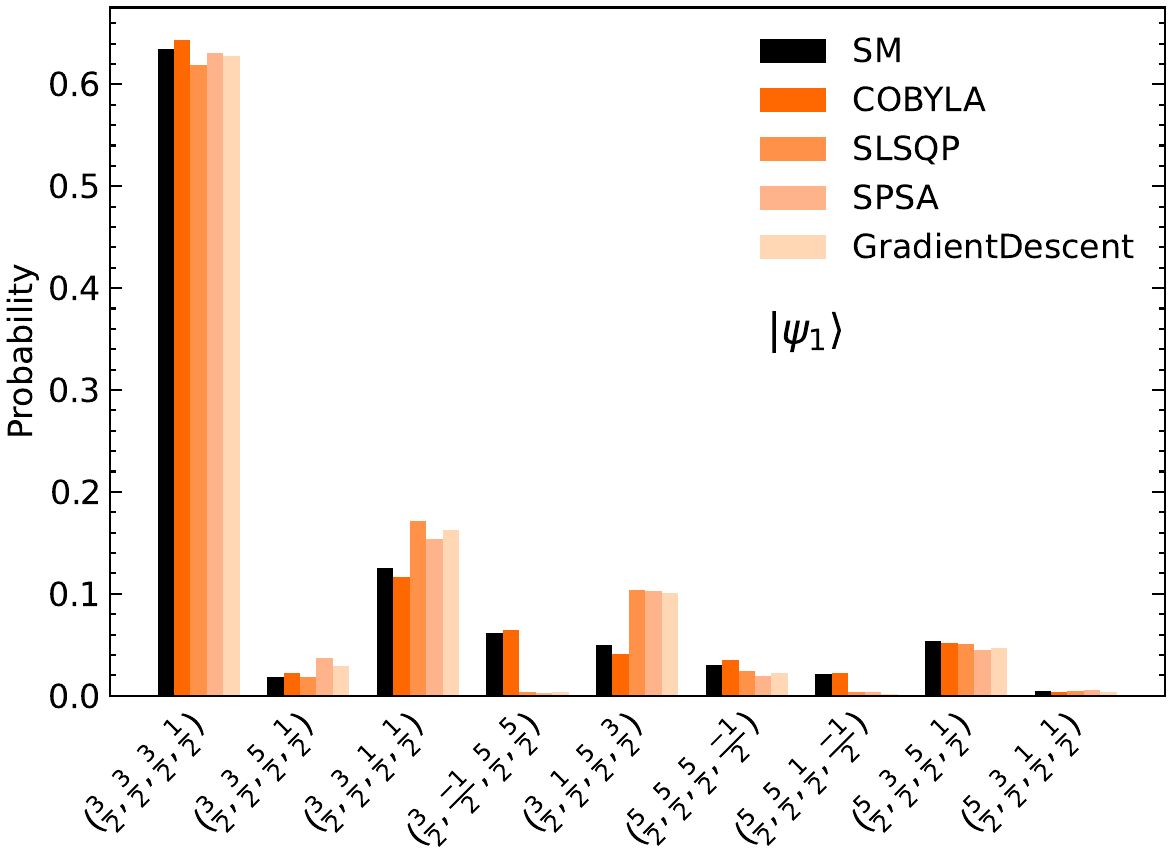}
    \caption{Components of the first excited state wave-function with different optimizers in comparison with the Shell model. The $x$-axis labels represent $(j_1,m_{\alpha_1},j_2,m_{\alpha_2})$ and $(m_{\alpha_1}+m_{\alpha_2}) = 2$.}
    \label{2p_wf}
\end{figure}

\begin{figure}[h]
\includegraphics[width=\textwidth,height=\textheight,keepaspectratio]{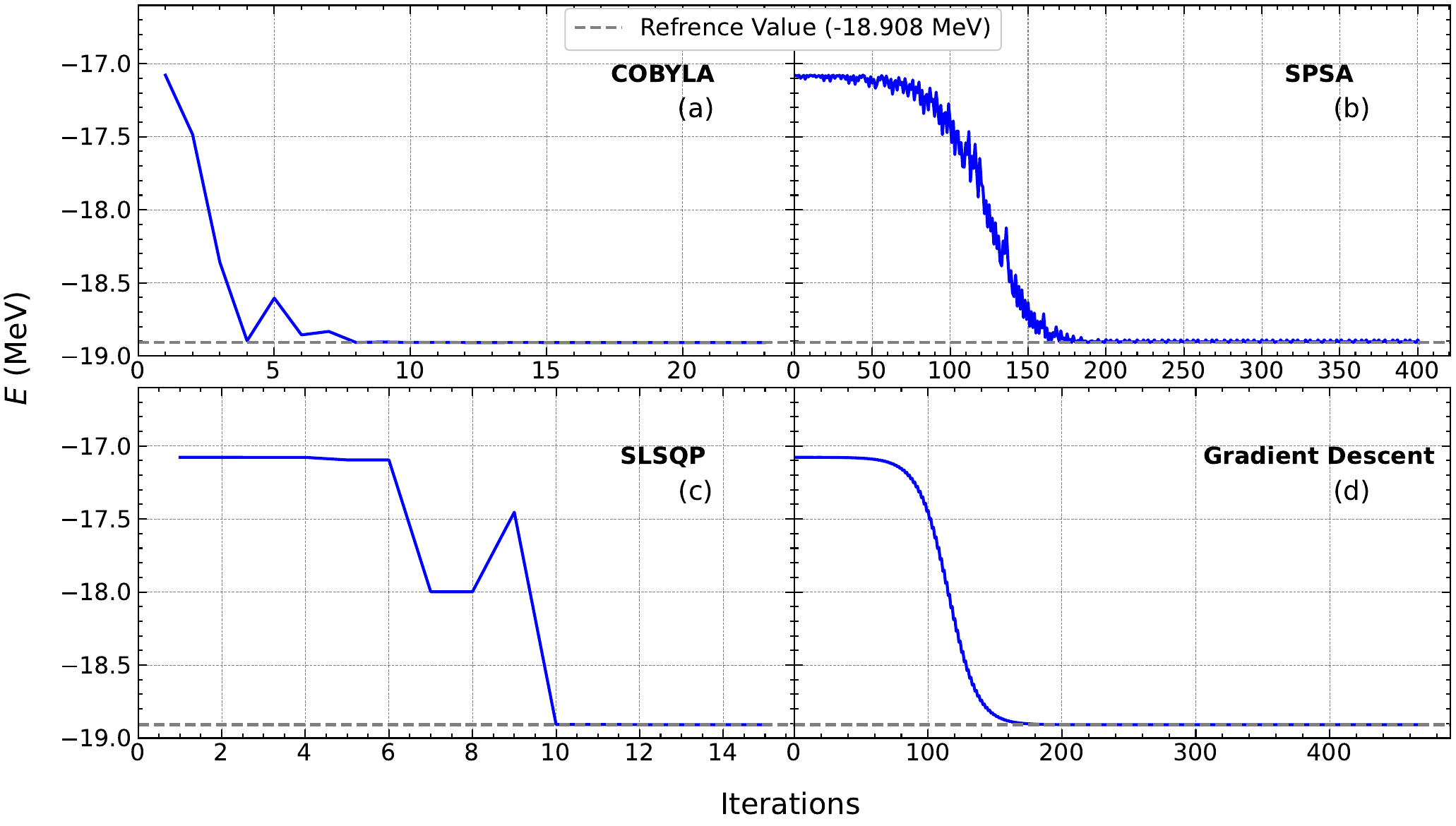}
   \caption{ Different optimizers convergence results for 12 Qubit $4^{+}$ ground state.}
  \label{12_excited_4p}
\end{figure}

\begin{figure}[h!]
    \centering    \includegraphics[width=0.65\textwidth,height=0.32\textheight]{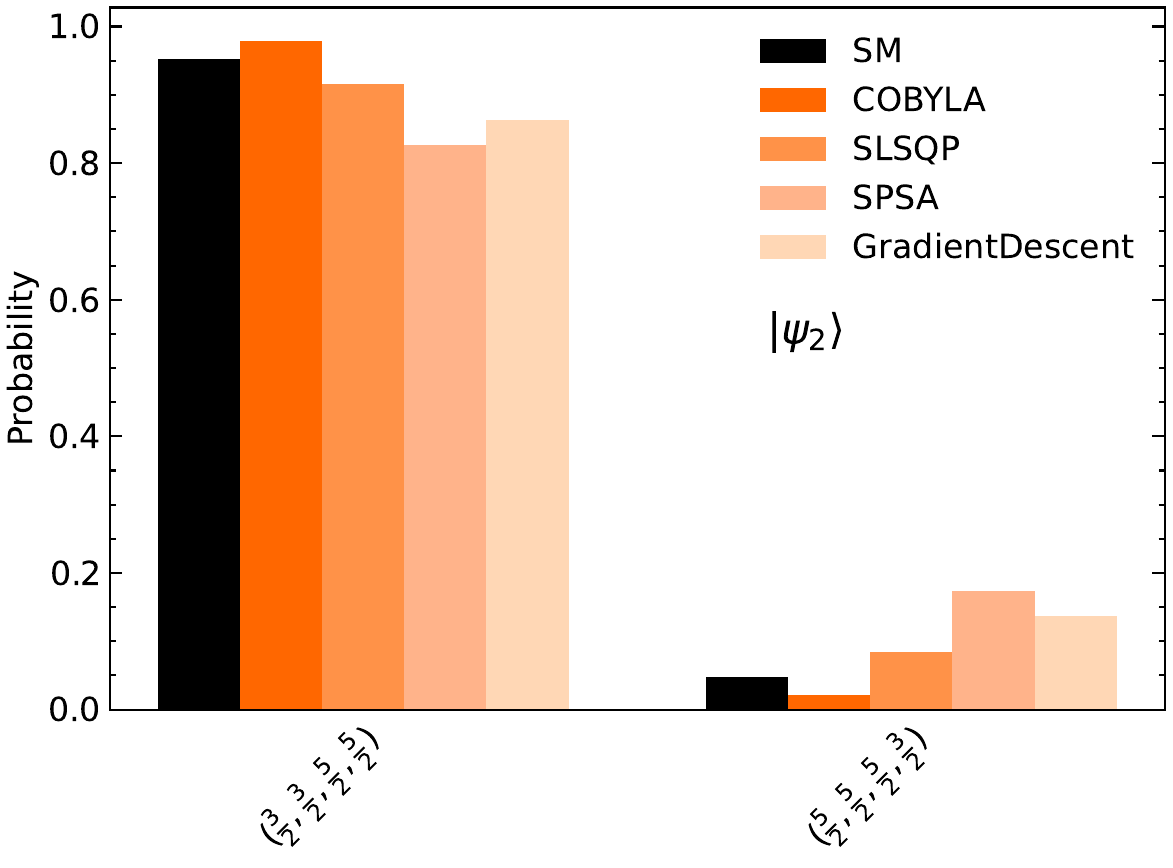}
    \caption{Components of the ground state wave-function with different optimizers in comparison with the Shell model.The $x$-axis labels represent $(j_1,m_{\alpha_1},j_2,m_{\alpha_2})$ and $(m_{\alpha_1}+m_{\alpha_2}) = 4$.}
    \label{4p_wf}
\end{figure}

Ground state calculations are performed using the ground state ansatz given and motivated previously.  In Figure \ref{12_ground}, we compare the different optimizer results discussed in Sec. \ref{subsec:optimizer} for the ground state. As illustrated in the figure, our prepared ansatz converges with all four optimizers. Notably, COBYLA converges fastest within 80 iterations, achieving an accuracy of $10^{-3}$ MeV shown in Table \ref{result}. The SLSQP also shows a similar convergence rate. This efficient convergence may be attributed to the problem-based ansatz preparation, which mitigates the presence of secondary minima. However, the other two optimizers are less advantageous for the prepared ansatz, requiring over 2000 iterations to converge to the desired state with a less accurate expectation value.
The comparison between the VQE wavefunction and the classical shell model wavefunction is shown in Figure \ref{G_wf}. The wavefunction is reproduced nicely with the VQE calculation.  As $^{58}$Ni is above magic nuclei $^{56}$Ni ($Z$ = 28, $N$=28), it is expected to exhibit single-particle nature as seen in the wavefunction for the ground state. Since the maximal occupied Qubit states represent the same orbital (in this case $1p_{3/2}$ with $> 80\% $ probability), this shows the single-particle nature of that wavefunction.



\subsection{First excited state}

In Figure \ref{12_excited_2p}, we compare the different optimizer results discussed in Sec. \ref{subsec:optimizer} for the first excited state. Similar to the ground state, our prepared ansatz converges with all four optimizers. Again, COBYLA exhibits the fastest convergence, achieving convergence in 105 iterations with an accuracy of $10^{-2}$ MeV, as detailed in Table \ref{result}. The remaining three optimizers converge close to the exact value but fall short of achieving precise accuracy. Similar to the ground state, SPSA and Gradient Descent take longer (more iterations) to converge. The comparison between the VQE wavefunction and the Shell model wavefunction is shown in Figure \ref{2p_wf}. The VQE calculation with COBYLA well-reproduces the reference shell model wavefunction, while the other optimizers are reasonable but miss e.g. a correct distribution across the $(3/2,-1/2,5/2,5/2)$ and $(3/2,1/2,5/2,-1/2)$ configurations.  In the m-scheme wavefunction, this first excited $J=2$ state is dominated by the $(1p_{3/2})^2$ and $(1p_{3/2}1p_{1/2})$ components, with probabilities of 63.49\% and 12.54\%, respectively. These characteristics are mirrored in the VQE wavefunction.


\subsection{Second excited state}

In Figure \ref{12_excited_4p}, we compare the optimizer results discussed in Sec. \ref{subsec:optimizer} for the second excited state. The prepared ansatz converges with all four optimizers to the exact reference value as detailed in Table \ref{result}. Notably, COBYLA and SLSQP converge in approximately 10 iterations. The perturbation parameter in SPSA has been tuned to 0.01 to get better convergence since it was showing oscillatory behaviour at the default calibration \cite{spsa2}. The comparison between the VQE wavefunction and the Shell model wavefunction is shown in Figure \ref{4p_wf}. The VQE calculation closely reproduces the dominant reference wavefunction, which is $1p_{3/2}0f_{5/2}$.

\begin{table}[h]
    \caption{Properties of ansatz used for the ground state, first and second excited state with the number of parameters, CNOT gates(2-qubit), RY gates(1-qubit), H gate(1-qubit), X gate(1-qubit) and circuit depth.}
    \label{Circ_prop}
    \centering
    \begin{tabular}{cccccc}
    \hline \hline
       State  & No. parameters & & 2-qubit & 1-qubit & Depth  \\
       \hline
        G.S. & 5 & & 70  & 72 & 96 \\
        $1^{st}$ e.s.& 7 && 82  & 78 & 108\\
        $2^{nd}$ e.s.& 1 && 4 & 4 & 7\\
    \hline \hline       
    \end{tabular}
\end{table}

\begin{table}[h]
    \caption{Summary of the results for the ground state (g.s.), first excited state ($1^{st}$ e.s.), and second excited state ($2^{nd}$ e.s.) alongside the shell model value for $^{58}$Ni in $pf$ model space. The exact result, obtained with exact diagonalization of the Hamiltonian and the $E_{UCC}$ energy is obtained by minimizing the Hamiltonian using ansatz.}
    \label{result}
    \centering
    \begin{tabular}{ccc|cccc}
    \hline \hline
         &  &  & \multicolumn{4}{c}{$E_{UCC}$}    \\
         ~~State~~ & ~~~SM~~ & ~~$E_{exact}$~~ & ~COBYLA~ & ~SLSQP~ & ~~SPSA~~ & ~~GD~ \\
       \hline
        G.S. & -20.990 & -20.990 & -20.990 & -20.990 & -20.988 & -20.990 \\
        $1^{st}$ e.s. & -20.024 & -20.024 & -20.022 & -19.919 & -19.912 & -19.918 \\
        $2^{nd}$ e.s. & -18.908 & -18.908 & -18.908 & -18.908 & -18.908 & -18.908  \\
    \hline \hline       
    \end{tabular}
\end{table}

\section{Conclusion}
\label{sec:conclusion}
The nuclear shell model provides a method to calculate properties of atomic nuclei in which nucleons interact in a given model space of single-particle states. By mapping single particle states to qubits, and using a suitable method to prepare wavefunctions -- in our case, tailored ansatz and VQE algorithm -- we have been able to extend the simulated quantum computation of nuclei up to the $^{58}$Ni isotope, finding exact results for the lowest $0^+$, $2^+$, and $4^+$ states.

Prospects for the method include a straightforward extension to other nuclei in the $fp$ shell, which requires no new qubits but more complicated circuits; possible simplification of the circuits through compilation or approximation techniques; and extension to heavier nuclei where the increasing single particle model space involves a linear increase in qubit number, automatically accompanied by an exponential increase in the size of the Hilbert space which hinders classical shell model calculations. Such simulated quantum computer calculations as we have shown will form the check of real quantum computation for future calculations.

\section*{Acknowledgements}
This work is supported by the UK STFC under the Quantum Technologies for Fundamental Physics programme, with grant ST/W006472/1.  We acknowledge useful discussions with Joe Gibbs about the wave function ansatzes.

\appendix{}
\section{$m$-scheme TBMEs} 
\label{appendix:A}

\begin{table}[h]
    \caption{The $m$-scheme TBMEs $\overline{v}_{\alpha\beta\gamma\delta}$ in MeV (Equation \ref{eq:m_scheme}) of JUN45 interaction \cite{jun45}. There is one to one mapping between the $m$-scheme basis ($\alpha$) and the qubits (Table \ref{q_mapping}).The core is taken as $^{56}$Ni, and the single-particle energies are considered -9.8280, -8.7087 and -7.8388 MeV for the $p_{3/2}$, $f_{5/2}$ and $p_{1/2}$ orbital, respectively. The TBMEs should be multiplied by a mass-scaling factor ${(A/58)}^{-0.3}$ in $J$-scheme, for any nuclei of mass $A$. Here $T$ = 1 and $M_T$ = $m_{t\alpha}+m_{t\beta}$, as we have calculated only for the same type of nuclei ($nn$ and $pp$). }
    \label{m_scheme_TBME}
    \centering
    \begin{ruledtabular}
    \begin{tabular}{cccccccccccccccccc}
         ~~$\alpha$~~ & ~~$\beta$~~ & ~~$\gamma$~~ & ~$\delta$~ & ~M~ & ~~$\overline{v}$ ~~ & ~~$\alpha$~~ & ~~$\beta$~~ & ~~$\gamma$~~ & ~$\delta$~ & ~M~ & ~~$\overline{v}$ ~~& ~~$\alpha$~~ & ~~$\beta$~~ & ~~$\gamma$~~ & ~$\delta$~ & ~M~ & ~~$\overline{v}$ ~~ \\
         \hline
   1	&  0	&  1	&  0	&	0  &   -0.2017  &  3	&  2	&  1	&  0	&	0  &   0.4476     &     0	&  7	&  1	&  6	&	0  &   -0.0815  \\
   1	&  0	&  1	&  6	&	0  &   -0.1637   &  3	&  2	&  1	&  6	&	0  &   -0.1637    &      0	&  7	&  3	&  2	&	0  &   -0.1637   \\
   1	&  0	&  3	&  2	&	0  &   0.4476    &  3	&  2	&  3	&  2	&	0  &   -0.2017   &      0	&  7	&  3	&  8	&	0  &   -0.0862    \\
   1	&  0	&  3	&  8	&	0  &   0.0668    &  3	&  2	&  3	&  8	&	0  &   0.0668     &      0	&  7	&  3	&  10	&	0  &   -0.1607    \\
   1	&  0	&  3	&  10	&	0  &   -0.2966   &  3	&  2	&  3	&  10	&	0  &   -0.2966    &      0	&  7	&  2	&  9	&	0  &   -0.1130       \\
   1	&  0	&  2	&  9	&	0  &   0.0668    &  3	&  2	&  2	&  9	&	0  &   0.0668     &      0	&  7	&  2	&  11	&	0  &   -0.1781     \\
   1	&  0	&  2	&  11	&	0  &   -0.2966   &  3	&  2	&  2	&  11	&	0  &   -0.2966    &      0	&  7	&  0	&  7	&	0  &   0.2850       \\
   1	&  0	&  0	&  7	&	0  &   -0.1637   &   3	&  2	&  0	&  7	&	0  &   -0.1637    &      0	&  7	&  5	&  4	&	0  &   -0.3415      \\
   1	&  0	&  5	&  4	&	0  &   -0.5585   &   3	&  2	&  5	&  4	&	0  &   0.1277     &      0	&  7	&  7	&  6	&	0  &   -0.1464      \\
   1	&  0	&  7	&  6	&	0  &   0.3000    &   3	&  2	&  7	&  6	&	0  &   -0.3862    &      0	&  7	&  9	&  8	&	0  &   0.1951       \\
   1	&  0	&  9	&  8	&	0  &   -0.1707   &   3	&  2	&  9	&  8	&	0  &   0.5154     &      0	&  7	&  9	&  10	&	0  &   -0.2178      \\
   1	&  0	&  9	&  10	&	0  &   -0.1156   &   3	&  2	&  9	&  10	&	0  &   -0.1156    &      0	&  7	&  8	&  11	&	0  &   0.0578          \\
   1	&  0	&  8	&  11	&	0  &   0.1156    &   3	&  2	&  8	&  11	&	0  &   0.1156     &       5	&  4	&  1	&  0	&	0  &   -0.5585    \\
   1	&  0	&  11	&  10	&	0  &   -0.8593   &   3	&  2	&  11	&  10	&	0  &   0.8593     &        5	&  4	&  1	&  6	&	0  &   -0.3415   \\
   1	&  6	&  1	&  0	&	0  &   -0.1637   &   3	&  8	&  1	&  0	&	0  &   0.0668     &        5	&  4	&  3	&  2	&	0  &   0.1277    \\
   1	&  6	&  1	&  6	&	0  &   0.2850  &   3	&  8	&  1	&  6	&	0  &   -0.1130    &        5	&  4	&  3	&  8	&	0  &   0.0597     \\
   1	&  6	&  3	&  2	&	0  &   -0.1637   &   3	&  8	&  3	&  2	&	0  &   0.0668     &        5	&  4	&  3	&  10	&	0  &   -0.2178    \\
   1	&  6	&  3	&  8	&	0  &   -0.1130   &    3	&  8	&  3	&  8	&	0  &   0.0873   &        5	&  4	&  2	&  9	&	0  &   0.0597     \\
   1	&  6	&  3	&  10	&	0  &   -0.1781   &      3	&  8	&  3	&  10	&	0  &   0.0798 &        5	&  4	&  2	&  11	&	0  &   -0.2178       \\
   1	&  6	&  2	&  9	&	0  &   -0.0862   &      3	&  8	&  2	&  9	&	0  &   -0.2902&        5	&  4	&  0	&  7	&	0  &   -0.3415        \\
   1	&  6	&  2	&  11	&	0  &   -0.1607   &      3	&  8	&  2	&  11	&	0  &   0.0585 &        5	&  4	&  5	&  4	&	0  &   -0.4098       \\
   1	&  6	&  0	&  7	&	0  &   -0.0815   &      3	&  8	&  0	&  7	&	0  &   -0.0862&        5	&  4	&  7	&  6	&	0  &   0.4424        \\
   1	&  6	&  5	&  4	&	0  &   -0.3415   &      3	&  8	&  5	&  4	&	0  &   0.0597 &        5	&  4	&  9	&  8	&	0  &   -0.3327       \\
   1	&  6	&  7	&  6	&	0  &   -0.1464   &      3	&  8	&  7	&  6	&	0  &   -0.1793&        5	&  4	&  9	&  10	&	0  &   -0.2071        \\
   1	&  6	&  9	&  8	&	0  &   0.1951    &      3	&  8	&  9	&  8	&	0  &   -0.2390&        5	&  4	&  8	&  11	&	0  &   0.2071       \\
   1	&  6	&  9	&  10	&	0  &   -0.0578   &      3	&  8	&  9	&  10	&	0  &   0.1216 &        5	&  4	&  11	&  10	&	0  &   -0.3401       \\
   1	&  6	&  8	&  11	&	0  &   0.2178    &      3	&  8	&  8	&  11	&	0  &   0.0091 &     7	&  6	&  1	&  0	&	0  &   0.3000       \\ 
   3	&  10	&  1	&  0	&	0  &   -0.2966   &     2	&  9	&  0	&  7	&	0  &   -0.1130 &      7	&  6	&  1	&  6	&	0  &   -0.1464           \\  
   3	&  10	&  1	&  6	&	0  &   -0.1781   &     2	&  9	&  5	&  4	&	0  &   0.0597 &       7	&  6	&  3	&  2	&	0  &   -0.3862        \\  
   3	&  10	&  3	&  2	&	0  &   -0.2966   &     2	&  9	&  7	&  6	&	0  &   -0.1793 &      7	&  6	&  3	&  8	&	0  &   -0.1793     \\  
   3	&  10	&  3	&  8	&	0  &   0.0798    &     2	&  9	&  9	&  8	&	0  &   -0.2390 &      7	&  6	&  3	&  10	&	0  &   -0.0436     \\ 
   3	&  10	&  3	&  10	&	0  &   0.1507  &     2	&  9	&  9	&  10	&	0  &   -0.0091 &      7	&  6	&  2	&  9	&	0  &   -0.1793     \\   
   3	&  10	&  2	&  9	&	0  &   0.0585    &     2	&  9	&  8	&  11	&	0  &   -0.1216 &      7	&  6	&  2	&  11	&	0  &   -0.0436     \\ 
   3	&  10	&  2	&  11	&	0  &   -0.3644   &     2	&  11	&  1	&  0	&	0  &   -0.2966 &      7	&  6	&  0	&  7	&	0  &   -0.1464     \\  
   3	&  10	&  0	&  7	&	0  &   -0.1607   &     2	&  11	&  1	&  6	&	0  &   -0.1607 &      7	&  6	&  5	&  4	&	0  &   0.4424     \\  
   3	&  10	&  5	&  4	&	0  &   -0.2178   &     2	&  11	&  3	&  2	&	0  &   -0.2966 &      7	&  6	&  7	&  6	&	0  &   -0.2343     \\  
   3	&  10	&  7	&  6	&	0  &   -0.0436   &     2	&  11	&  3	&  8	&	0  &   0.0585  &      7	&  6	&  9	&  8	&	0  &   0.5082    \\  
   3	&  10	&  9	&  8	&	0  &   0.1742    &     2	&  11	&  3	&  10	&	0  &   -0.3644 &      7	&  6	&  9	&  10	&	0  &   -0.0414   \\
   3	&  10	&  9	&  10	&	0  &   -0.1220   &     2	&  11	&  2	&  9	&	0  &   0.0798  &      7	&  6	&  8	&  11	&	0  &   0.0414   \\
   3	&  10	&  8	&  11	&	0  &   0.1220    &     2	&  11	&  2	&  11	&	0  &   0.1507&      7	&  6	&  11	&  10	&	0  &   0.3401    \\
   2	&  9	&  1	&  0	&	0  &   0.0668    &     2	&  11	&  0	&  7	&	0  &   -0.1781 &      9	&  8	&  1	&  0	&	0  &   -0.1707  \\
   2	&  9	&  1	&  6	&	0  &   -0.0862   &     2	&  11	&  5	&  4	&	0  &   -0.2178 &       9	&  8	&  1	&  6	&	0  &   0.1951   \\
   2	&  9	&  3	&  2	&	0  &   0.0668    &     2	&  11	&  7	&  6	&	0  &   -0.0436 &       9	&  8	&  3	&  2	&	0  &   0.5154  \\
   2	&  9	&  3	&  8	&	0  &   -0.2902   &     2	&  11	&  9	&  8	&	0  &   0.1742  &       9	&  8	&  3	&  8	&	0  &   -0.2390  \\
   2	&  9	&  3	&  10	&	0  &   0.0585    &     2	&  11	&  9	&  10	&	0  &   -0.1220 &       9	&  8	&  3	&  10	&	0  &   0.1742  \\
   2	&  9	&  2	&  9	&	0  &   0.0873  &     2	&  11	&  8	&  11	&	0  &   0.1220  &       9	&  8	&  2	&  9	&	0  &   -0.2390   \\
   2	&  9	&  2	&  11	&	0  &   0.0798    &     0	&  7	&  1	&  0	&	0  &   -0.1637 &       9	&  8	&  2	&  11	&	0  &   0.1742     \\

    \end{tabular}
    \end{ruledtabular}
\end{table}
  
 \addtocounter{table}{-1} 
 
 \begin{table}
    \leavevmode
    \caption{{Continuation.}}
    \begin{ruledtabular}
\begin{tabular}{cccccccccccccccccc} 

 ~~$\alpha$~~ & ~~$\beta$~~ & ~~$\gamma$~~ & ~$\delta$~ & ~M~ & ~~$\overline{v}$ ~~ & ~~$\alpha$~~ & ~~$\beta$~~ & ~~$\gamma$~~ & ~$\delta$~ & ~M~ & ~~$\overline{v}$ ~~& ~~$\alpha$~~ & ~~$\beta$~~ & ~~$\gamma$~~ & ~$\delta$~ & ~M~ & ~~$\overline{v}$ ~~ \\
         \hline
 
   9	&  8	&  0	&  7	&	0  &   0.1951     &    1	&  8	& 1	&  10	& 1 &	-0.1091  &        2 & 7	& 9	& 11 & 1  &		-0.0813             \\
   9	&  8	&  5	&  4	&	0  &   -0.3327    &      1	&  8	& 3	&  9	& 1 &	-0.1807  &        0	& 5	& 1	& 2	 & 1  &	-0.2113              \\
   9	&  8	&  7	&  6	&	0  &   0.5082     &      1	&  8	& 3	&  11	& 1 &	-0.1766  &        0	& 5	& 1	& 8	 & 1  &	-0.0645             \\
   9	&  8	&  9	&  8	&	0  &   -0.3440   &      1	&  8	& 2	&  7	& 1 &	-0.1361  &        0	& 5	& 1	& 10 & 1  &		-0.0924               \\
   9	&  8	&  9	&  10	&	0  &   0.1657     &      1	&  8	& 0	&  5	& 1 &	-0.0645  &        0	& 5	& 3	& 9	 & 1  &	-0.0270             \\
   9	&  8	&  8	&  11	&	0  &   -0.1657    &      1	&  8	& 5	&  6	& 1 &	-0.3857  &        0	& 5	& 3	& 11 & 1  &		-0.1991              \\
   9	&  8	&  11	&  10	&	0  &   -0.3401    &      1	&  8	& 7	&  8	& 1 &	0.0487   &        0	& 5	& 2	& 7	 & 1  &	-0.0702             \\
   9	&  10	&  1	&  0	&	0  &   -0.1156    &      1	&  8	& 7	&  10	& 1 &	-0.0578  &        0	& 5	& 0	& 5	 & 1  &	0.3394              \\
   9	&  10	&  1	&  6	&	0  &   -0.0578    &      1	&  8	& 9	&  11	& 1 &	0.2106   &        0	& 5	& 5	& 6	 & 1  &	-0.3415             \\
   9	&  10	&  3	&  2	&	0  &   -0.1156    &      1	&  10	& 1	&  2	& 1 &	-0.2966  &        0	& 5	& 7	& 8	 & 1  &	0.1542              \\
   9	&  10	&  3	&  8	&	0  &   0.1216     &      1	& 10	& 1	& 8	& 1	&  -0.1091   &       0	& 5	& 7	& 10 & 1 &		-0.1874          \\
   9	&  10	&  3	&  10	&	0  &   -0.1220    &       1	& 10	& 1	& 10& 1	&  	0.3328 &       0	& 5	& 9	& 11 & 1 &		0.0431             \\
   9	&  10	&  2	&  9	&	0  &   -0.0091    &       1	& 10	& 3	& 9	& 1	&  0.1091    &       5	& 6	& 1	& 2	 & 1 &	-0.3338          \\
   9	&  10	&  2	&  11	&	0  &   -0.1220    &       1	& 10	& 3	& 11& 1	&  	-0.3155  &       5	& 6	& 1	& 8	 & 1 &	-0.3857            \\
   9	&  10	&  0	&  7	&	0  &   -0.2178    &       1	& 10	& 2	& 7	& 1	&  -0.0413   &       5	& 6	& 1	& 10 & 1 &		-0.1687           \\
   9	&  10	&  5	&  4	&	0  &   -0.2071    &       1	& 10	& 0	& 5	& 1	&  -0.0924   &       5	& 6	& 3	& 9	 & 1 &	0.1335           \\
   9	&  10	&  7	&  6	&	0  &   -0.0414    &       1	& 10	& 5	& 6	& 1	&  -0.1687   &       5	& 6	& 3	& 11 & 1 &		-0.2922           \\
   9	&  10	&  9	&  8	&	0  &   0.1657     &       1	& 10	& 7	& 8	& 1	&  0.1067    &       5	& 6	& 2	& 7	 & 1 &	-0.1890         \\
   9	&  10	&  9	&  10	&	0  &   -0.0024   &       1	& 10	& 7	& 10& 1	&  	-0.0996  &       5	& 6	& 0	& 5	 & 1 &	-0.3415             \\
   9	&  10	&  8	&  11	&	0  &   0.4076     &       1	& 10	& 9	& 11& 1	&  	0.0704   &       5	& 6	& 5	& 6	 & 1 &	0.0326          \\
   8	&  11	&  1	&  0	&	0  &   0.1156     &       3 & 9	& 1	&  2	& 1  &0.1929     &       5	& 6	& 7	&  8  & 1 & 0.1387          \\
   8	&  11	&  1	&  6	&	0  &   0.2178     &       3	& 9	& 1	&  8	& 1  &-0.1807    &       5	& 6	& 7	&  1  & 1 & 	-0.2619           \\
   8	&  11	&  3	&  2	&	0  &   0.1156     &       3	& 9	& 1	&  10   & 1  &0.1091    &        5	& 6	& 9	&  1  & 1 & 	0.1852          \\
   8	&  11	&  3	&  8	&	0  &   0.0091     &       3	& 9	& 3	&  9	& 1  &-0.0018   &       7	& 8	& 1	&  2  & 1 & 0.2111            \\
   8	&  11	&  3	&  10	&	0  &   0.1220     &       3	& 9	& 3	&  11   & 1  &	0.1675    &      7	& 8	& 1	&  8  & 1 & 0.0487            \\
   8	&  11	&  2	&  9	&	0  &   -0.1216    &       3	& 9	& 2	&  7	& 1  &-0.2448    &       7	& 8	& 1	&  1  & 1 & 	0.1067            \\
   8	&  11	&  2	&  11	&	0  &   0.1220     &       3	& 9	& 0	&  5	& 1  &-0.0270    &       7	& 8	& 3	&  9  & 1 & -0.4226           \\
   8	&  11	&  0	&  7	&	0  &   0.0578     &       3	& 9	& 5	&  6	& 1  &0.1335     &       7	& 8	& 3	&  1  & 1 & 	0.1848          \\
   8	&  11	&  5	&  4	&	0  &   0.2071     &       3	& 9	& 7	&  8	& 1  &-0.4226    &       7	& 8	& 2	&  7  & 1 & -0.1196           \\
   8	&  11	&  7	&  6	&	0  &   0.0414     &       3	& 9	& 7	&  10   & 1  &0.1480    &        7	& 8	& 0	&  5  & 1 & 0.1542          \\
   8	&  11	&  9	&  8	&	0  &   -0.1657    &       3 	& 9	& 9	& 11	& 1	&-0.0720 &       7	& 8	   &   5	& 6	  & 1 &0.1387             \\
   8	&  11	&  9	&  10	&	0  &   0.4076     &        3	& 11& 	1& 	2	& 1	&-0.5137 &        7	& 8	   &   7	& 8	  & 1 & 0.1643            \\
   8	&  11	&  8	&  11	&	0  &   -0.0024   &        3	& 11& 	1& 	8	& 1	&-0.1766 &        7	& 8	   &   7	& 10  & 1 &	0.1657             \\
   11	&  10	&  1	&  0	&	0  &   -0.8593    &        3	& 11& 	1& 	10  & 1	&-0.3155 &        7	& 8	   &   9	& 11  & 1 &	-0.1171             \\
   11	&  10	&  3	&  2	&	0  &   0.8593     &        3	& 11& 	3& 	9	& 1  &	0.1675 &      7	& 10   &   	1	& 2	  & 1 &-0.1888              \\
   11	&  10	&  5	&  4	&	0  &   -0.3401    &        3	& 11& 	3& 	11  & 1	&-0.0315 &       7	& 10   &   	1	& 8	  & 1 &-0.0578              \\
   11	&  10	&  7	&  6	&	0  &   0.3401     &        3	& 11& 	2& 	7	& 1	&-0.0413 &        7	& 10   &   	1	& 10  & 1 &	-0.0996            \\
   11	&  10	&  9	&  8	&	0  &   -0.3401    &        3	& 11& 	0& 	5	& 1	&-0.1991 &        7	& 10   &   	3	& 9	  & 1 &0.1480             \\
   11	&  10	&  11	&  10	&	0  &   0.0309   &        3	& 11& 	5& 	6	& 1	&-0.2922 &        7	& 10   &   	3	& 11  & 1 &	-0.1725              \\
   1	&  2	&  1	&  2	& 1 &	0.2459      &        3	& 11& 	7& 	8	& 1 &0.1848 &         7	& 10   &   	2	& 7	  & 1 &-0.1137       \\ 
   1	&  2	&  1	&  8	& 1 &	-0.2005       &      3	&  11	& 7	& 10& 1 &	-0.1725  &        7	& 10   &   	0	& 5	  & 1 &-0.1874           \\ 
   1	&  2	&  1	&  10	& 1 &	-0.2966       &      3	&  11	& 9	& 11& 1 &	0.1220   &        7	& 10   &   	5	& 6	  & 1 &-0.2619           \\ 
   1	&  2	&  3	&  9	& 1 &	0.1929        &      2	&  7	& 1	& 2	& 1 &-0.0546   &          7	& 10   &   	7	& 8	  & 1 &0.1657      \\ 
   1	&  2	&  3	&  11	& 1 &	-0.5137       &      2	&  7	& 1	& 8	& 1 &-0.1361   &          7	& 10   &   	7	& 10  & 1 &	-0.1383         \\ 
   1	&  2	&  2	&  7	& 1 &	-0.0546       &      2	&  7	& 1	& 10& 1 &	-0.0413  &        7	& 10   &   	9	& 11  & 1 &	0.3842            \\ 
   1	&  2	&  0	&  5	& 1 &	-0.2113       &      2	&  7	& 3	& 9	& 1 &-0.2448   &          9	& 11   &   	1	& 2	  & 1 &0.1335          \\ 
   1	&  2	&  5	&  6	& 1 &	-0.3338       &      2	&  7	& 3	& 11& 1 &	-0.0413  &        9	& 11   &   	1	& 8	  & 1 &0.2106            \\ 
   1	&  2	&  7	&  8	& 1 &	0.2111        &      2	&  7	& 2	& 7	& 1 & 0.1912  &          9	& 11   &   	1	& 10  & 1 &	0.0704           \\ 
   1	&  2	&  7	&  10	& 1 &	-0.1888       &      2	&  7	& 0	& 5	& 1 &-0.0702   &          9	& 11   &   	3	& 9	  & 1 &-0.0720          \\ 
   1	&  2	&  9	&  11	& 1 &	0.1335        &      2	&  7	& 5	& 6	& 1 &-0.1890   &          9	& 11   &   	3	& 11  & 1 &	0.1220             \\
   1	&  8	&  1	&  2	& 1 &	-0.2005       &      2	&  7	& 7	& 8	& 1 &-0.1196   &          9  &	11	&2	& 7	& 1 &	-0.0813           \\
   1	&  8	&  1	&  8	& 1 &	0.2159      &      2	&  7	& 7	& 10& 1 &	-0.1137  &         9	&11	&0	& 5	& 1 &	0.0431          \\

    \end{tabular}
    \end{ruledtabular}
\end{table}
  
 \addtocounter{table}{-1} 
 
 \begin{table}
    \leavevmode
    \caption{{Continuation.}}
    \begin{ruledtabular}
\begin{tabular}{cccccccccccccccccc} 

 ~~$\alpha$~~ & ~~$\beta$~~ & ~~$\gamma$~~ & ~$\delta$~ & ~M~ & ~~$\overline{v}$ ~~ & ~~$\alpha$~~ & ~~$\beta$~~ & ~~$\gamma$~~ & ~$\delta$~ & ~M~ & ~~$\overline{v}$ ~~& ~~$\alpha$~~ & ~~$\beta$~~ & ~~$\gamma$~~ & ~$\delta$~ & ~M~ & ~~$\overline{v}$ ~~ \\
         \hline 
   9	& 11	& 5	& 6	& 1 &	0.1852             & 5	&  8	& 3	& 7	 &	2 &   -0.0001        &   1	& 4	& 3	& 0	 & -1 &	-0.2113                     \\      
   9	& 11	& 7	& 8	& 1 &	-0.1171            & 5	&  8	& 2	& 5	 &	2 &   -0.3585        &    1	& 4	& 3	& 6	 & -1 &	-0.0702                    \\
   9	& 11	& 7	& 10& 1 &	0.3842             & 5	&  8	& 5	& 8	 &	2 &   0.1423       &    1	& 4	& 2	& 8	 & -1 &	-0.0270                    \\
   9	& 11	& 9	& 11& 1 &	0.1334           & 5	&  8	& 5	& 10 &	2 &   	-0.2071      &    1	& 4	& 2	& 10 & -1 &		-0.1991                       \\ 
   1	& 3	&  1	&  3	&2  & 0.2459         & 5	&  8	& 7	& 9	 &	2 &   0.1471         &    1	& 4	& 0	& 9	 & -1 &	-0.0645                     \\              
   1	& 3	&  1	&  9	&2  & -0.1336          & 5	&  8	& 7	& 11 &	2 &   	0.0926       &    1	& 4	& 0	& 11 & -1 &		-0.0924                     \\
   1	& 3	&  1	&  11	&2    & -0.5932        & 5	&  10	& 1	& 3	 &	2 &   -0.2111        &    1	& 4	& 7	& 4	 & -1 &	-0.3415                      \\
   1	& 3	&  3	&  7	&2  & 0.2182           & 5	&  10	& 1	& 9	 &	2 &   -0.0431        &    1	& 4	& 9	& 6	 & -1 &	0.1542                   \\
   1	& 3	&  2	&  5	&2  & -0.2440          & 5	&  10	& 1	& 11 &	2 &   	-0.2227      &    1	& 4	& 8	& 10 & -1 &		-0.0431                      \\
   1	& 3	&  5	&  8	&2  & -0.2360          & 5	&  10	& 3	& 7	 &	2 &   0.1434         &    1	& 4	& 6	& 11 & -1 &		0.1874                   \\
   1	& 3	&  5	&  10	&2    & -0.2111        & 5	&  10	& 2	& 5	 &	2 &   -0.2420        &    3	& 0	& 1	& 4	 & -1 &	-0.2113                    \\
   1	& 3	&  7	&  9	&2  & 0.3166           & 5	&  10	& 5	& 8	 &	2 &   -0.2071        &      3	& 0	& 3	& 0	 &-1 & 0.2459                 \\
   1	& 3	&  7	&  11	&2    & 0.0944         & 5	&  10	& 5	& 10 &	2 &   	-0.2742     &      3	& 0	& 3	& 6	 &-1 & -0.0546                      \\
   1	& 9	&  1	&  3	&2  & -0.1336          & 5	&  10	& 7	& 9	 &	2 &   0.2778         &      3	& 0	& 2	& 8	 &-1 & 0.1929                 \\
   1	& 9	   &   1	&  9	& 2 &	0.1115   & 5	&  10	& 7	& 11 &	2 &   	0.3038       &      3	& 0	& 2	& 10 &-1 & 	-0.5137                            \\
   1	& 9	   &   1	&  11	& 2 &	-0.1383    & 7	&  9	& 1	& 3	 &	2 &   0.3166         &      3	& 0	& 0	& 9	 &-1 & -0.2005                         \\
   1	& 9	   &   3	&  7	& 2 &	-0.2611    & 7	&  9	& 1	& 9	 &	2 &   0.0487         &      3	& 0	& 0	& 11 &-1 & 	-0.2966                         \\
   1	& 9	   &   2	&  5	& 2 &	-0.1386    & 7	&  9	& 1	& 11 &	2 &   	0.3201       &      3	& 0	& 7	& 4	 &-1 & -0.3338                           \\
   1	& 9	   &   5	&  8	& 2 &	-0.3273    & 7	&  9	& 3	& 7	 &	2 &   -0.4780        &      3	& 0	& 9	& 6	 &-1 & 0.2111                          \\
   1	& 9	   &   5	&  10	& 2 &	-0.0431    & 7	&  9	& 2	& 5	 &	2 &   0.2672         &      3	& 0	& 8	& 10 &-1 & 	-0.1335                         \\
   1	& 9	   &   7	&  9	& 2 &	0.0487     & 7	&  9	& 5	& 8	 &	2 &   0.1471         &      3	& 0	& 6	& 11 &-1 & 	0.1888                      \\
   1	& 9	   &   7	&  11	& 2 &	0.1793     & 7	&  9	& 5	& 10 &	2 &   	0.2778       &        3	& 6	& 1	& 4	 & -1 & -0.0702                        \\
   1	& 11   &   	1	&  3	& 2 &	-0.5932    & 7	&  9	& 7	& 9	 &	2 &   0.0546       &        3	& 6	& 3	& 0	 & -1 & -0.0546                         \\
   1	& 11   &   	1	&  9	& 2 &	-0.1383    & 7	&  9	& 7	& 11 &	2 &   	-0.1243      &        3	& 6	& 3	& 6	 & -1 & 0.1912                          \\
   1	& 11   &   	1	&  11	& 2 &	-0.2137   & 7	&  11	& 1	& 3	 &	2 &   0.0944         &        3	& 6	& 2	& 8	 & -1 & -0.2448                        \\
   1	& 11   &   	3	&  7	& 2 &	0.2258     & 7	&  11	& 1	& 9	 &	2 &   0.1793         &        3	& 6	& 2	& 10 & -1 & 	-0.0413                      \\
   1	& 11   &   	2	&  5	& 2 &	-0.2525    & 7	&  11	& 1	& 11 &	2 &   	0.0996       &        3	& 6	& 0	& 9	 & -1 & -0.1361                         \\
   1	& 11   &   	5	&  8	& 2 &	-0.2386    & 7	&  11	& 3	& 7	 &	2 &   -0.1295        &        3	& 6	& 0	& 11 & -1 & 	-0.0413                        \\
   1	& 11   &   	5	&  10	& 2 &	-0.2227    & 7	&  11	& 2	& 5	 &	2 &   -0.0379        &        3	& 6	& 7	& 4	 & -1 & -0.1890                        \\
   1	& 11   &   	7	&  9	& 2 &	0.3201     & 7	&  11	& 5	& 8	 &	2 &   0.0926         &        3	& 6	& 9	& 6	 & -1 & -0.1196                      \\
   1	& 11   &   	7	&  11	& 2 &	0.0996     & 7	&  11	& 5	& 10 &	2 &   	0.3038       &        3	& 6	& 8	& 10 & -1 & 	0.0813                      \\
   3	& 7	   &   1	&  3	& 2 &	0.2182     & 7	&  11	& 7	& 9	 &	2 &   -0.1243        &          3	& 6	   &   6	& 11 &	-1 & 	0.1137                     \\
   3	& 7	   &   1	&  9	& 2 &	-0.2611    & 7	&  11	& 7	& 11 &	2 &   	0.2693     &          2	& 8	   &   1	& 4	 &	-1 & -0.0270                         \\
   3	& 7	   &   1	&  11	& 2 &	0.2258     & 1	& 7	   &   1	&  7   & 3    & -0.0484 &          2	& 8	   &   3	& 0	 &	-1 & 0.1929                    \\
   3	& 7	   &   3	&  7	& 2 &	-0.0150   &   1 & 7	 &   3	&  5   & 3    & -0.3045  &          2	& 8	   &   3	& 6	 &	-1 & -0.2448                    \\
   3	& 7	   &   2	&  5	& 2 &	-0.1252    &   1& 7	   &   5	&  9   & 3    & -0.3086  &          2	& 8	   &   2	& 8	 &	-1 & -0.0018                   \\
   3	& 7	   &   5	&  8	& 2 &	-0.0001    &   1& 7	   &   5	&  11   & 3    & 0.1265   &          2	& 8	   &   2	& 10 &	-1 & 	0.1675                  \\
   3	& 7	   &   5	&  10	& 2 &	0.1434     &   3& 5	   &   1	&  7   & 3    & -0.3045  &          2	& 8	   &   0	& 9	 &	-1 & -0.1807                  \\
   3	& 7	   &   7	&  9	& 2 &	-0.4780    &   3& 5	   &   3	&  5   & 3    & 0.1089 &          2	& 8	   &   0	& 11 &	-1 & 	0.1091                    \\
   3	& 7	   &   7	&  11	& 2 &	-0.1295    &   3& 5	   &   5	&  9   & 3    & -0.2391  &          2	& 8	   &   7	& 4	 &	-1 & 0.1335                   \\
   2	& 5	   &   1	&  3	& 2 &	-0.2440    &   3& 5	   &   5	&  11   & 3    & -0.1633  &         2	& 8	   &   9	& 6	 &	-1 & -0.4226                   \\
   2	& 5	   &   1	&  9	& 2 &	-0.1386    &   5& 9	   &   1	&  7   & 3    & -0.3086  &          2	& 8	   &   8	& 10 &	-1 & 	0.0720                   \\
   2	& 5	   &   1	&  11	& 2 &	-0.2525    &   5& 9	   &   3	&  5   & 3    & -0.2391  &          2	& 8	   &   6	& 11 &	-1 & 	-0.1480                   \\
   2	& 5	   &   3	&  7	& 2 &	-0.1252    &   5& 9	   &   5	&  9   & 3    & 0.2520 &          2	& 10   &   	1	& 4	 &	-1 & -0.1991                    \\
   2	& 5	   &   2	&  5	& 2 &	0.2488   &   5& 11   &   	1	&  7   & 3    & 0.1265   &          2	& 10   &   	3	& 0	 &	-1 & -0.5137                   \\
   2	& 5	   &   5	&  8	& 2 &	-0.3585    &   5& 11   &   	3	&  5   & 3    & -0.1633  &          2	& 10   &   	3	& 6	 &	-1 & -0.0413                   \\
   2	& 5	   &   5	&  10	& 2 &	-0.2420    &   5& 11   &   	5	&  11   & 3    & 0.4051 &          2	& 10   &   	2	& 8	 &	-1 & 0.1675                   \\
   2	& 5	   &   7	&  9	& 2 &	0.2672     &    1	& 5	& 1	& 5	& 4 &	-0.2842         &          2	& 10   &   	2	& 10 &	-1 & 	-0.0315                    \\
   2	& 5	   &   7	&  11	& 2 &	-0.0379    &    1	& 5	& 5	& 7	& 4 &	-0.3904          &          2	& 10   &   	0	& 9	 &	-1 & -0.1766                    \\
   5	& 8	   &   1	&  3	& 2 &	-0.2360    &    5	& 7	& 1	& 5	& 4 &	-0.3904          &          2	& 10   &   	0	& 11 &	-1 & 	-0.3155                    \\
   5	& 8	   &   1	&  9	& 2 &	-0.3273    &    5	& 7	& 5	& 7	& 4 &	0.2520         &          2	& 10   &   	7	& 4	 &	-1 & -0.2922                   \\
   5	& 8	   &   1	&  11	& 2 &	-0.2386    &   1	& 4	& 1	&  4 & -1 &	0.3394         &          2	& 10   &   	9	& 6	 &	-1 & 0.1848                 \\

       \end{tabular}
    \end{ruledtabular}
\end{table}
  
 \addtocounter{table}{-1} 
 
 \begin{table}
    \leavevmode
    \caption{{Continuation.}}
    \begin{ruledtabular}
\begin{tabular}{cccccccccccccccccc} 

 ~~$\alpha$~~ & ~~$\beta$~~ & ~~$\gamma$~~ & ~$\delta$~ & ~M~ & ~~$\overline{v}$ ~~ & ~~$\alpha$~~ & ~~$\beta$~~ & ~~$\gamma$~~ & ~$\delta$~ & ~M~ & ~~$\overline{v}$ ~~& ~~$\alpha$~~ & ~~$\beta$~~ & ~~$\gamma$~~ & ~$\delta$~ & ~M~ & ~~$\overline{v}$ ~~ \\
         \hline 
   2	&  10	&  8	&  10	&  -1 &	-0.1220  &   8	&  10	&  9	&  6	& -1 &	0.1171   &        0	& 10   &   	0&	8	& -2 &	-0.1383               \\  
   2	&  10	&  6	&  11	&  -1 &	0.1725 &    8	&  10	&  8	&  10	& -1 &	0.1334   &       0	& 10   &   	2&	6	& -2 &	0.2258                 \\
   0	&  9	&  1	&  4	&  -1 &	-0.0645  &    8	&  10	&  6	&  11	& -1 &	0.3842   &          3	& 4	& 3	& 4	& -2 & 0.2488             \\
   0	&  9	&  3	&  0	&  -1 &	-0.2005  &    6	&  11	&  1	&  4	& -1 &	0.1874   &          3	& 4	& 2	& 0	& -2 & -0.2440             \\
   0	&  9	&  3	&  6	&  -1 &	-0.1361  &    6	&  11	&  3	&  0	& -1 &	0.1888   &           0	& 10   &   	0	& 10 & -2  & 	-0.2137             \\
   0	&  9	&  2	&  8	&  -1 &	-0.1807  &    6	&  11	&  3	&  6	& -1 &	0.1137   &            0	& 10   &   	9	& 4	 & -2  & -0.2386             \\
   0	&  9	&  2	&  10	&  -1 &	-0.1766  &    6	&  11	&  2	&  8	& -1 &	-0.1480   &           0	& 10   &   	8	& 6	 & -2  & 0.3201              \\
   0	&  9	&  0	&  9	&  -1 &	0.2159 &    6	&  11	&  2	&  10	& -1 &	0.1725   &            0	& 10   &   	6	& 10 & -2  & 	-0.0996             \\ 
   0	&  9	&  0	&  11	&  -1 &	-0.1091  &    6	&  11	&  0	&  9	& -1 &	0.0578   &            0	& 10   &   	4	& 11 & -2  & 	0.2227             \\
   0	&  9	&  7	&  4	&  -1 &	-0.3857  &    6	&  11	&  0	&  11	& -1 &	0.0996   &            9	& 4	   &   3	& 4	 & -2  & -0.3585             \\
   0	&  9	&  9	&  6	&  -1 &	0.0487  &    6	&  11	&  7	&  4	& -1 &	0.2619   &            9	& 4	   &   2	& 0	 & -2  & -0.2360             \\
   0	&  9	&  8	&  10	&  -1 &	-0.2106  &    6	&  11	&  9	&  6	& -1 &	-0.1657   &           9	& 4	   &   2	& 6	 & -2  & -0.0001              \\
   0	&  9	&  6	&  11	&  -1 &	0.0578  &    6	&  11	&  8	&  10	& -1 &	0.3842   &            9	& 4	   &   0	& 8	 & -2  & -0.3273             \\
   0	&  11	&  1	&  4	&  -1 &	-0.0924  &    6	&  11	&  6	&  11	& -1 &	-0.1383   &          9	& 4	   &   0	& 10 & -2  & 	-0.2386               \\
   0	&  11	&  3	&  0	&  -1 &	-0.2966    &    8	&  10	&  0	&  11	&  -1 &	-0.0704 &       9	& 4	   &   9	& 4	 & -2  & 0.1423                          \\
   0	&  11	&  3	&  6	&  -1 &	-0.0413    &     8	&  10	&  7	&  4	&  -1 &	-0.1852   &       9	& 4	   &   8	& 6	 & -2 &	0.1471                      \\
   0	&  11	&  2	&  8	&  -1 &	0.1091     &      3	& 4	& 2	& 6	& -2 & -0.1252  &                     9	& 4	   &   6	& 10 & -2 &		-0.0926           \\
   0	&  11	&  2	&  10	&  -1 &	-0.3155    &      3	& 4	& 0	& 8	& -2 & -0.1386  &                     9	& 4	   &   4	& 11 & -2 &		0.2071            \\
   0	&  11	&  0	&  9	&  -1 &	-0.1091    &      3	& 4	& 0	& 10& -2 & 	-0.2525 &                     8	& 6	   &   3	& 4	 & -2 &	0.2672             \\
   0	&  11	&  0	&  11	&  -1 &	0.3328   &      3	& 4	& 9	& 4	& -2 & -0.3585  &                     8	& 6	   &   2	& 0	 & -2 &	0.3166            \\ 
   0	&  11	&  7	&  4	&  -1 &	-0.1687    &      3	& 4	& 8	& 6	& -2 & 0.2672   &                     8	& 6	   &   2	& 6	 & -2 &	-0.4780           \\
   0	&  11	&  9	&  6	&  -1 &	0.1067     &      3	& 4	& 6	& 10& -2 & 	0.0379  &                     8	& 6	   &   0	& 8	 & -2 &	0.0487           \\
   0	&  11	&  8	&  10	&  -1 &	-0.0704    &      3	& 4	& 4	& 11& -2 & 	0.2420  &                     8	& 6	   &   0	& 10 & -2 &		0.3201            \\
   0	&  11	&  6	&  11	&  -1 &	0.0996     &      2	& 0	& 3	& 4	& -2 & -0.2440  &                     8	& 6	   &   9	& 4	 & -2 &	0.1471           \\
   7	&  4	&  1	&  4	&  -1 &	-0.3415    &       2	& 0	& 2	& 0	 & -2 & 0.2459 &               8	& 6	   &   8	& 6	 & -2 &	0.0546               \\
   7	&  4	&  3	&  0	&  -1 &	-0.3338    &         2	& 0	& 2	& 6	 & -2 & 0.2182   &                8	& 6	   &   6	& 10 & -2 &		0.1243             \\
   7	&  4	&  3	&  6	&  -1 &	-0.1890    &         2	& 0	& 0	& 8	 & -2 & -0.1336  &                8	& 6	   &   4	& 11 & -2 &		-0.2778              \\
   7	&  4	&  2	&  8	&  -1 &	0.1335     &         2	& 0	& 0	& 10 & -2 & 	-0.5932 &             6	& 10   &   	3	& 4	 & -2 &	0.0379                 \\
   7	&  4	&  2	&  10	&  -1 &	-0.2922    &         2	& 0	& 9	& 4	 & -2 & -0.2360  &                6	& 10   &   	2	& 0	 & -2 &	-0.0944              \\
   7	&  4	&  0	&  9	&  -1 &	-0.3857    &         2	& 0	& 8	& 6	 & -2 & 0.3166   &                6	& 10   &   	2	& 6	 & -2 &	0.1295             \\
   7	&  4	&  0	&  11	&  -1 &	-0.1687    &         2	& 0	& 6	& 10 & -2 & 	-0.0944 &             6	& 10   &   	0	& 8	 & -2 &	-0.1793                  \\
   7	&  4	&  7	&  4	&  -1 &	0.0326   &         2	& 0	& 4	& 11 & -2 & 	0.2111  &             6	& 10   &   	0	& 10 & -2 &		-0.0996                 \\ 
   7	&  4	&  9	&  6	&  -1 &	0.1387     &         2	& 6	& 3	& 4	 & -2 & -0.1252  &                6	& 10   &   	9	& 4	 & -2 &	-0.0926             \\
   7	&  4	&  8	&  10	&  -1 &	-0.1852    &         2	& 6	& 2	& 0	 & -2 & 0.2182   &                6	& 10   &   	8	& 6	 & -2 &	0.1243             \\
   7	&  4	&  6	&  11	&  -1 &	0.2619     &          2	& 6	& 2	& 6	 & -2 & -0.0150 &              6	& 10	& 6	& 10 & -2 & 		0.2693            \\
   9	&  6	&  1	&  4	&  -1 &	0.1542     &       2	& 6	& 0	& 8	 & -2 &	-0.2611  &               6	& 10	& 4	& 11 & -2 & 		0.3038           \\
   9	&  6	&  3	&  0	&  -1 &	0.2111     &       2	& 6	& 0	& 10 & -2 &		0.2258  &            4	& 11	& 3	& 4	 & -2 & 	0.2420              \\
   9	&  6	&  3	&  6	&  -1 &	-0.1196    &       2	& 6	& 9	& 4	 & -2 &	-0.0001  &               4	& 11	& 2	& 0	 & -2 & 	0.2111            \\
   9	&  6	&  2	&  8	&  -1 &	-0.4226    &       2	& 6	& 8	& 6	 & -2 &	-0.4780  &               4	& 11	& 2	& 6	 & -2 & 	-0.1434            \\
   9	&  6	&  2	&  10	&  -1 &	0.1848     &       2	& 6	& 6	& 10 & -2 &		0.1295  &            4	& 11	& 0	& 8	 & -2 & 	0.0431              \\
   9	&  6	&  0	&  9	&  -1 &	0.0487     &       2	& 6	& 4	& 11 & -2 &		-0.1434 &            4	& 11	& 0	& 10 & -2 & 		0.2227               \\
   9	&  6	&  0	&  11	&  -1 &	0.1067     &       0	& 8	& 3	& 4	 & -2 &	-0.1386  &               4	& 11	& 9	& 4	 & -2 & 	0.2071           \\
   9	&  6	&  7	&  4	&  -1 &	0.1387     &       0	& 8	& 2	& 0	 & -2 &	-0.1336  &               4	& 11	& 8	& 6	 & -2 & 	-0.2778           \\
   9	&  6	&  9	&  6	&  -1 &	0.1643   &       0	& 8	& 2	& 6	 & -2 &	-0.2611  &               4	& 11	& 6	& 10 & -2 & 		0.3038            \\ 
   9	&  6	&  8	&  10	&  -1 &	0.1171     &        0	& 8	& 0	& 8	 & -2 &	0.1115 &               4	& 11	& 4	& 11 & -2 & 		-0.2742         \\
   9	&  6	&  6	&  11	&  -1 &	-0.1657    &      0	& 8	   &   0	&10	& -2 &	-0.1383 &             2	   & 4	& 2	& 4	 & -3 &	0.1089            \\
   8	&  10	&  1	&  4	&  -1 &	-0.0431    &      0	& 8	   &   9	&4	& -2 &	-0.3273 &              2	& 4	& 0	& 6	 & -3 &	-0.3045            \\
   8	&  10	&  3	&  0	&  -1 &	-0.1335    &      0	& 8	   &   8	&6	& -2 &	0.0487  &              2	& 4	& 8	& 4	 & -3 &	-0.2391           \\
   8	&  10	&  3	&  6	&  -1 &	0.0813     &      0	& 8	   &   6	&10	& -2 &	-0.1793 &              2	& 4	& 4	& 10 & -3 &		0.1633           \\
   8	&  10	&  2	&  8	&  -1 &	0.0720     &      0	& 8	   &   4	&11	& -2 &	0.0431  &              0	& 6	& 2	& 4	 & -3 &	-0.3045          \\
   8	&  10	&  2	&  10	&  -1 &	-0.1220    &      0	& 10   &   	3&	4	& -2 &	-0.2525 &              0	& 6	& 0	& 6	 & -3 &	-0.0484            \\
   8	&  10	&  0	&  9	&  -1 &	-0.2106    &      0	& 10   &   	2&	0	& -2 &	-0.5932 &              0	& 6	& 8	& 4	 & -3 &	-0.3086            \\

       \end{tabular}
    \end{ruledtabular}
\end{table}
  
 \addtocounter{table}{-1} 
 
 \begin{table}
    \leavevmode
    \caption{{Continuation.}}
    \begin{ruledtabular}
\begin{tabular}{cccccccccccccccccc} 

 ~~$\alpha$~~ & ~~$\beta$~~ & ~~$\gamma$~~ & ~$\delta$~ & ~M~ & ~~$\overline{v}$ ~~ & ~~$\alpha$~~ & ~~$\beta$~~ & ~~$\gamma$~~ & ~$\delta$~ & ~M~ & ~~$\overline{v}$ ~~& ~~$\alpha$~~ & ~~$\beta$~~ & ~~$\gamma$~~ & ~$\delta$~ & ~M~ & ~~$\overline{v}$ ~~ \\

 \hline

   0	&  6	& 4	& 10 & -3 &	-0.1265  &     4	&  10	& 2	& 4	 & -3 &	0.1633  &      0	& 4	& 6	& 4	& -4 &	-0.3904   \\
   8	&  4	& 2	& 4	 & -3 &	-0.2391 &   4	&  10	& 0	& 6	 & -3 &	-0.1265  &    6	& 4	& 0	& 4	& -4 &	-0.3904   \\
   8	&  4	& 0	& 6	 & -3 &	-0.3086 &   4	&  10	& 4	& 10 & -3 &		0.4051  &    6	& 4	& 6	& 4	& -4 &	0.2520   \\
   8	&  4	& 8	& 4	 & -3 &	0.2520 &    0  & 4	& 0	& 4	& -4 &	-0.2842  &  &&&&& \\

    \end{tabular}
    \end{ruledtabular}
\end{table}

\clearpage
\bibliographystyle{apsrev4-2}
\bibliography{Bibliography}

\begin{thebibliography}{49}%
\makeatletter
\providecommand \@ifxundefined [1]{%
 \@ifx{#1\undefined}
}%
\providecommand \@ifnum [1]{%
 \ifnum #1\expandafter \@firstoftwo
 \else \expandafter \@secondoftwo
 \fi
}%
\providecommand \@ifx [1]{%
 \ifx #1\expandafter \@firstoftwo
 \else \expandafter \@secondoftwo
 \fi
}%
\providecommand \natexlab [1]{#1}%
\providecommand \enquote  [1]{``#1''}%
\providecommand \bibnamefont  [1]{#1}%
\providecommand \bibfnamefont [1]{#1}%
\providecommand \citenamefont [1]{#1}%
\providecommand \href@noop [0]{\@secondoftwo}%
\providecommand \href [0]{\begingroup \@sanitize@url \@href}%
\providecommand \@href[1]{\@@startlink{#1}\@@href}%
\providecommand \@@href[1]{\endgroup#1\@@endlink}%
\providecommand \@sanitize@url [0]{\catcode `\\12\catcode `\$12\catcode `\&12\catcode `\#12\catcode `\^12\catcode `\_12\catcode `\%12\relax}%
\providecommand \@@startlink[1]{}%
\providecommand \@@endlink[0]{}%
\providecommand \url  [0]{\begingroup\@sanitize@url \@url }%
\providecommand \@url [1]{\endgroup\@href {#1}{\urlprefix }}%
\providecommand \urlprefix  [0]{URL }%
\providecommand \Eprint [0]{\href }%
\providecommand \doibase [0]{https://doi.org/}%
\providecommand \selectlanguage [0]{\@gobble}%
\providecommand \bibinfo  [0]{\@secondoftwo}%
\providecommand \bibfield  [0]{\@secondoftwo}%
\providecommand \translation [1]{[#1]}%
\providecommand \BibitemOpen [0]{}%
\providecommand \bibitemStop [0]{}%
\providecommand \bibitemNoStop [0]{.\EOS\space}%
\providecommand \EOS [0]{\spacefactor3000\relax}%
\providecommand \BibitemShut  [1]{\csname bibitem#1\endcsname}%
\let\auto@bib@innerbib\@empty
\bibitem [{\citenamefont {Vi{\v s}{\v n}{\'a}k}(2015)}]{visnak_quantum_2015}%
  \BibitemOpen
  \bibfield  {author} {\bibinfo {author} {\bibfnamefont {J.}~\bibnamefont {Vi{\v s}{\v n}{\'a}k}},\ }\href {https://doi.org/10.1051/epjconf/201510001008} {\bibfield  {journal} {\bibinfo  {journal} {EPJ Web of Conferences}\ }\textbf {\bibinfo {volume} {100}},\ \bibinfo {pages} {01008} (\bibinfo {year} {2015})}\BibitemShut {NoStop}%
\bibitem [{\citenamefont {Vi{\v s}{\v n}{\'a}k}\ and\ \citenamefont {Vesel{\'y}}(2017)}]{visnak_quantum_2017}%
  \BibitemOpen
  \bibfield  {author} {\bibinfo {author} {\bibfnamefont {J.}~\bibnamefont {Vi{\v s}{\v n}{\'a}k}}\ and\ \bibinfo {author} {\bibfnamefont {P.}~\bibnamefont {Vesel{\'y}}},\ }\href {https://doi.org/10.1051/epjconf/201715401030} {\bibfield  {journal} {\bibinfo  {journal} {EPJ Web of Conferences}\ }\textbf {\bibinfo {volume} {154}},\ \bibinfo {pages} {01030} (\bibinfo {year} {2017})}\BibitemShut {NoStop}%
\bibitem [{\citenamefont {Dumitrescu}\ \emph {et~al.}(2018)\citenamefont {Dumitrescu}, \citenamefont {McCaskey}, \citenamefont {Hagen}, \citenamefont {Jansen}, \citenamefont {Morris}, \citenamefont {Papenbrock}, \citenamefont {Pooser}, \citenamefont {Dean},\ and\ \citenamefont {Lougovski}}]{dumitrescu_cloud_2018}%
  \BibitemOpen
  \bibfield  {author} {\bibinfo {author} {\bibfnamefont {E.~F.}\ \bibnamefont {Dumitrescu}}, \bibinfo {author} {\bibfnamefont {A.~J.}\ \bibnamefont {McCaskey}}, \bibinfo {author} {\bibfnamefont {G.}~\bibnamefont {Hagen}}, \bibinfo {author} {\bibfnamefont {G.~R.}\ \bibnamefont {Jansen}}, \bibinfo {author} {\bibfnamefont {T.~D.}\ \bibnamefont {Morris}}, \bibinfo {author} {\bibfnamefont {T.}~\bibnamefont {Papenbrock}}, \bibinfo {author} {\bibfnamefont {R.~C.}\ \bibnamefont {Pooser}}, \bibinfo {author} {\bibfnamefont {D.~J.}\ \bibnamefont {Dean}},\ and\ \bibinfo {author} {\bibfnamefont {P.}~\bibnamefont {Lougovski}},\ }\href {https://doi.org/10.1103/PhysRevLett.120.210501} {\bibfield  {journal} {\bibinfo  {journal} {Physical Review Letters}\ }\textbf {\bibinfo {volume} {120}},\ \bibinfo {pages} {210501} (\bibinfo {year} {2018})},\ \bibinfo {note} {publisher: American Physical Society}\BibitemShut {NoStop}%
\bibitem [{\citenamefont {Gibbs}\ \emph {et~al.}(2024)\citenamefont {Gibbs}, \citenamefont {Holmes},\ and\ \citenamefont {Stevenson}}]{gibbs_exploiting_2024}%
  \BibitemOpen
  \bibfield  {author} {\bibinfo {author} {\bibfnamefont {J.}~\bibnamefont {Gibbs}}, \bibinfo {author} {\bibfnamefont {Z.}~\bibnamefont {Holmes}},\ and\ \bibinfo {author} {\bibfnamefont {P.}~\bibnamefont {Stevenson}},\ }\href {https://doi.org/10.48550/arXiv.2402.10277} {\bibinfo {title} {Exploiting symmetries in nuclear {Hamiltonians} for ground state preparation}} (\bibinfo {year} {2024}),\ \bibinfo {note} {arXiv:2402.10277}\BibitemShut {NoStop}%
\bibitem [{\citenamefont {Illa}\ \emph {et~al.}(2023)\citenamefont {Illa}, \citenamefont {Robin},\ and\ \citenamefont {Savage}}]{illa_quantum_2023}%
  \BibitemOpen
  \bibfield  {author} {\bibinfo {author} {\bibfnamefont {M.}~\bibnamefont {Illa}}, \bibinfo {author} {\bibfnamefont {C.~E.~P.}\ \bibnamefont {Robin}},\ and\ \bibinfo {author} {\bibfnamefont {M.~J.}\ \bibnamefont {Savage}},\ }\href {https://doi.org/10.1103/PhysRevC.108.064306} {\bibfield  {journal} {\bibinfo  {journal} {Physical Review C}\ }\textbf {\bibinfo {volume} {108}},\ \bibinfo {pages} {064306} (\bibinfo {year} {2023})}\BibitemShut {NoStop}%
\bibitem [{\citenamefont {Robin}\ and\ \citenamefont {Savage}(2023)}]{robin_quantum_2023}%
  \BibitemOpen
  \bibfield  {author} {\bibinfo {author} {\bibfnamefont {C.~E.~P.}\ \bibnamefont {Robin}}\ and\ \bibinfo {author} {\bibfnamefont {M.~J.}\ \bibnamefont {Savage}},\ }\href {https://doi.org/10.1103/PhysRevC.108.024313} {\bibfield  {journal} {\bibinfo  {journal} {Physical Review C}\ }\textbf {\bibinfo {volume} {108}},\ \bibinfo {pages} {024313} (\bibinfo {year} {2023})}\BibitemShut {NoStop}%
\bibitem [{\citenamefont {Roggero}\ \emph {et~al.}(2020)\citenamefont {Roggero}, \citenamefont {Gu}, \citenamefont {Baroni},\ and\ \citenamefont {Papenbrock}}]{roggero_preparation_2020}%
  \BibitemOpen
  \bibfield  {author} {\bibinfo {author} {\bibfnamefont {A.}~\bibnamefont {Roggero}}, \bibinfo {author} {\bibfnamefont {C.}~\bibnamefont {Gu}}, \bibinfo {author} {\bibfnamefont {A.}~\bibnamefont {Baroni}},\ and\ \bibinfo {author} {\bibfnamefont {T.}~\bibnamefont {Papenbrock}},\ }\href {https://doi.org/10.1103/PhysRevC.102.064624} {\bibfield  {journal} {\bibinfo  {journal} {Physical Review C}\ }\textbf {\bibinfo {volume} {102}},\ \bibinfo {pages} {1} (\bibinfo {year} {2020})}\BibitemShut {NoStop}%
\bibitem [{\citenamefont {Stevenson}(2023)}]{stevenson_p_d_comments_2023}%
  \BibitemOpen
  \bibfield  {author} {\bibinfo {author} {\bibfnamefont {P.~D.}\ \bibnamefont {Stevenson}},\ }\href {https://www.oldcitypublishing.com/journals/ijuc-home/ijuc-issue-contents/ijuc-volume-18-number-1-2023/ijuc-18-1-p-83-92/} {\bibfield  {journal} {\bibinfo  {journal} {International Journal of Unconventional Computing}\ }\textbf {\bibinfo {volume} {18}},\ \bibinfo {pages} {83} (\bibinfo {year} {2023})}\BibitemShut {NoStop}%
\bibitem [{\citenamefont {Zhang}\ \emph {et~al.}(2021)\citenamefont {Zhang}, \citenamefont {Xing}, \citenamefont {Yan}, \citenamefont {Wang},\ and\ \citenamefont {Zhu}}]{zhang_selected_2021}%
  \BibitemOpen
  \bibfield  {author} {\bibinfo {author} {\bibfnamefont {D.-B.}\ \bibnamefont {Zhang}}, \bibinfo {author} {\bibfnamefont {H.}~\bibnamefont {Xing}}, \bibinfo {author} {\bibfnamefont {H.}~\bibnamefont {Yan}}, \bibinfo {author} {\bibfnamefont {E.}~\bibnamefont {Wang}},\ and\ \bibinfo {author} {\bibfnamefont {S.-L.}\ \bibnamefont {Zhu}},\ }\href {https://doi.org/10.1088/1674-1056/abd761} {\bibfield  {journal} {\bibinfo  {journal} {Chinese Physics B}\ }\textbf {\bibinfo {volume} {30}},\ \bibinfo {pages} {020306} (\bibinfo {year} {2021})}\BibitemShut {NoStop}%
\bibitem [{\citenamefont {Stetcu}\ \emph {et~al.}(2022{\natexlab{a}})\citenamefont {Stetcu}, \citenamefont {Baroni},\ and\ \citenamefont {Carlson}}]{stetcu_variational_2021}%
  \BibitemOpen
  \bibfield  {author} {\bibinfo {author} {\bibfnamefont {I.}~\bibnamefont {Stetcu}}, \bibinfo {author} {\bibfnamefont {A.}~\bibnamefont {Baroni}},\ and\ \bibinfo {author} {\bibfnamefont {J.}~\bibnamefont {Carlson}},\ }\href {https://journals.aps.org/prc/abstract/10.1103/PhysRevC.105.064308} {\bibfield  {journal} {\bibinfo  {journal} {Physical Review C}\ }\textbf {\bibinfo {volume} {105}},\ \bibinfo {pages} {064308} (\bibinfo {year} {2022}{\natexlab{a}})}\BibitemShut {NoStop}%
\bibitem [{\citenamefont {Siwach}\ and\ \citenamefont {Arumugam}(2021{\natexlab{a}})}]{siwach_quantum_2021}%
  \BibitemOpen
  \bibfield  {author} {\bibinfo {author} {\bibfnamefont {P.}~\bibnamefont {Siwach}}\ and\ \bibinfo {author} {\bibfnamefont {P.}~\bibnamefont {Arumugam}},\ }\href {https://doi.org/10.1103/PhysRevC.104.034301} {\bibfield  {journal} {\bibinfo  {journal} {Physical Review C}\ }\textbf {\bibinfo {volume} {104}},\ \bibinfo {pages} {034301} (\bibinfo {year} {2021}{\natexlab{a}})}\BibitemShut {NoStop}%
\bibitem [{\citenamefont {Robbins}\ and\ \citenamefont {Love}(2021)}]{robbins_benchmarking_2021}%
  \BibitemOpen
  \bibfield  {author} {\bibinfo {author} {\bibfnamefont {K.}~\bibnamefont {Robbins}}\ and\ \bibinfo {author} {\bibfnamefont {P.~J.}\ \bibnamefont {Love}},\ }\href {https://journals.aps.org/pra/abstract/10.1103/PhysRevA.104.022412} {\bibfield  {journal} {\bibinfo  {journal} {Physical Review A}\ }\textbf {\bibinfo {volume} {104}},\ \bibinfo {pages} {022412} (\bibinfo {year} {2021})}\BibitemShut {NoStop}%
\bibitem [{\citenamefont {Chikaoka}\ and\ \citenamefont {Liang}(2022)}]{chikaoka_quantum_2022}%
  \BibitemOpen
  \bibfield  {author} {\bibinfo {author} {\bibfnamefont {A.}~\bibnamefont {Chikaoka}}\ and\ \bibinfo {author} {\bibfnamefont {H.}~\bibnamefont {Liang}},\ }\href {https://doi.org/10.1088/1674-1137/ac380a} {\bibfield  {journal} {\bibinfo  {journal} {Chinese Physics C}\ }\textbf {\bibinfo {volume} {46}},\ \bibinfo {pages} {024106} (\bibinfo {year} {2022})}\BibitemShut {NoStop}%
\bibitem [{\citenamefont {Li}\ \emph {et~al.}(2024{\natexlab{a}})\citenamefont {Li}, \citenamefont {Al-Khalili},\ and\ \citenamefont {Stevenson}}]{li_solving_2024}%
  \BibitemOpen
  \bibfield  {author} {\bibinfo {author} {\bibfnamefont {Y.~H.}\ \bibnamefont {Li}}, \bibinfo {author} {\bibfnamefont {J.}~\bibnamefont {Al-Khalili}},\ and\ \bibinfo {author} {\bibfnamefont {P.}~\bibnamefont {Stevenson}},\ }\href {https://doi.org/10.48550/arXiv.2402.01623} {\bibinfo {title} {Solving coupled {Non}-linear {Schr\"odinger} {Equations} via {Quantum} {Imaginary} {Time} {Evolution}}} (\bibinfo {year} {2024}{\natexlab{a}}),\ \bibinfo {note} {arXiv:2402.01623}\BibitemShut {NoStop}%
\bibitem [{\citenamefont {Li}\ \emph {et~al.}(2024{\natexlab{b}})\citenamefont {Li}, \citenamefont {Al-Khalili},\ and\ \citenamefont {Stevenson}}]{li_quantum_2024}%
  \BibitemOpen
  \bibfield  {author} {\bibinfo {author} {\bibfnamefont {Y.~H.}\ \bibnamefont {Li}}, \bibinfo {author} {\bibfnamefont {J.}~\bibnamefont {Al-Khalili}},\ and\ \bibinfo {author} {\bibfnamefont {P.}~\bibnamefont {Stevenson}},\ }\href {https://doi.org/10.48550/arXiv.2308.15425} {\bibinfo {title} {A {Quantum} {Simulation} {Approach} to {Implementing} {Nuclear} {Density} {Functional} {Theory} via {Imaginary} {Time} {Evolution}}} (\bibinfo {year} {2024}{\natexlab{b}}),\ \bibinfo {note} {arXiv:2308.15425}\BibitemShut {NoStop}%
\bibitem [{\citenamefont {Cervia}\ \emph {et~al.}(2021)\citenamefont {Cervia}, \citenamefont {Balantekin}, \citenamefont {Coppersmith}, \citenamefont {Johnson}, \citenamefont {Love}, \citenamefont {Poole}, \citenamefont {Robbins},\ and\ \citenamefont {Saffman}}]{cervia_lipkin_2021}%
  \BibitemOpen
  \bibfield  {author} {\bibinfo {author} {\bibfnamefont {M.~J.}\ \bibnamefont {Cervia}}, \bibinfo {author} {\bibfnamefont {A.~B.}\ \bibnamefont {Balantekin}}, \bibinfo {author} {\bibfnamefont {S.~N.}\ \bibnamefont {Coppersmith}}, \bibinfo {author} {\bibfnamefont {C.~W.}\ \bibnamefont {Johnson}}, \bibinfo {author} {\bibfnamefont {P.~J.}\ \bibnamefont {Love}}, \bibinfo {author} {\bibfnamefont {C.}~\bibnamefont {Poole}}, \bibinfo {author} {\bibfnamefont {K.}~\bibnamefont {Robbins}},\ and\ \bibinfo {author} {\bibfnamefont {M.}~\bibnamefont {Saffman}},\ }\href {https://doi.org/10.1103/PhysRevC.104.024305} {\bibfield  {journal} {\bibinfo  {journal} {Physical Review C}\ }\textbf {\bibinfo {volume} {104}},\ \bibinfo {pages} {024305} (\bibinfo {year} {2021})}\BibitemShut {NoStop}%
\bibitem [{\citenamefont {Brown}\ \emph {et~al.}(2010)\citenamefont {Brown}, \citenamefont {Munro},\ and\ \citenamefont {Kendon}}]{brown_using_2010}%
  \BibitemOpen
  \bibfield  {author} {\bibinfo {author} {\bibfnamefont {K.~L.}\ \bibnamefont {Brown}}, \bibinfo {author} {\bibfnamefont {W.~J.}\ \bibnamefont {Munro}},\ and\ \bibinfo {author} {\bibfnamefont {V.~M.}\ \bibnamefont {Kendon}},\ }\href {https://doi.org/10.3390/e12112268} {\bibfield  {journal} {\bibinfo  {journal} {Entropy}\ }\textbf {\bibinfo {volume} {12}},\ \bibinfo {pages} {2268} (\bibinfo {year} {2010})}\BibitemShut {NoStop}%
\bibitem [{\citenamefont {McArdle}\ \emph {et~al.}(2020)\citenamefont {McArdle}, \citenamefont {Endo}, \citenamefont {Aspuru-Guzik}, \citenamefont {Benjamin},\ and\ \citenamefont {Yuan}}]{RevModPhys.92.015003}%
  \BibitemOpen
  \bibfield  {author} {\bibinfo {author} {\bibfnamefont {S.}~\bibnamefont {McArdle}}, \bibinfo {author} {\bibfnamefont {S.}~\bibnamefont {Endo}}, \bibinfo {author} {\bibfnamefont {A.}~\bibnamefont {Aspuru-Guzik}}, \bibinfo {author} {\bibfnamefont {S.~C.}\ \bibnamefont {Benjamin}},\ and\ \bibinfo {author} {\bibfnamefont {X.}~\bibnamefont {Yuan}},\ }\href {https://doi.org/10.1103/RevModPhys.92.015003} {\bibfield  {journal} {\bibinfo  {journal} {Rev. Mod. Phys.}\ }\textbf {\bibinfo {volume} {92}},\ \bibinfo {pages} {015003} (\bibinfo {year} {2020})}\BibitemShut {NoStop}%
\bibitem [{\citenamefont {Ma}\ \emph {et~al.}(2020)\citenamefont {Ma}, \citenamefont {Govoni},\ and\ \citenamefont {Galli}}]{ma_quantum_2020}%
  \BibitemOpen
  \bibfield  {author} {\bibinfo {author} {\bibfnamefont {H.}~\bibnamefont {Ma}}, \bibinfo {author} {\bibfnamefont {M.}~\bibnamefont {Govoni}},\ and\ \bibinfo {author} {\bibfnamefont {G.}~\bibnamefont {Galli}},\ }\href {https://doi.org/10.1038/s41524-020-00353-z} {\bibfield  {journal} {\bibinfo  {journal} {npj Computational Materials}\ }\textbf {\bibinfo {volume} {6}},\ \bibinfo {pages} {1} (\bibinfo {year} {2020})}\BibitemShut {NoStop}%
\bibitem [{\citenamefont {McGrory}\ and\ \citenamefont {Wildenthal}(1980)}]{mcgrory_large-scale_1980}%
  \BibitemOpen
  \bibfield  {author} {\bibinfo {author} {\bibfnamefont {J.~B.}\ \bibnamefont {McGrory}}\ and\ \bibinfo {author} {\bibfnamefont {B.~H.}\ \bibnamefont {Wildenthal}},\ }\href@noop {} {\bibfield  {journal} {\bibinfo  {journal} {Annual Review of Nuclear and Particle Science}\ }\textbf {\bibinfo {volume} {30}},\ \bibinfo {pages} {383} (\bibinfo {year} {1980})}\BibitemShut {NoStop}%
\bibitem [{\citenamefont {Caurier}\ \emph {et~al.}(2005)\citenamefont {Caurier}, \citenamefont {Martínez-Pinedo}, \citenamefont {Nowacki}, \citenamefont {Poves},\ and\ \citenamefont {Zuker}}]{caurier_shell_2005}%
  \BibitemOpen
  \bibfield  {author} {\bibinfo {author} {\bibfnamefont {E.}~\bibnamefont {Caurier}}, \bibinfo {author} {\bibfnamefont {G.}~\bibnamefont {Martínez-Pinedo}}, \bibinfo {author} {\bibfnamefont {F.}~\bibnamefont {Nowacki}}, \bibinfo {author} {\bibfnamefont {A.}~\bibnamefont {Poves}},\ and\ \bibinfo {author} {\bibfnamefont {A.~P.}\ \bibnamefont {Zuker}},\ }\href {https://doi.org/10.1103/RevModPhys.77.427} {\bibfield  {journal} {\bibinfo  {journal} {Reviews of Modern Physics}\ }\textbf {\bibinfo {volume} {77}},\ \bibinfo {pages} {427} (\bibinfo {year} {2005})}\BibitemShut {NoStop}%
\bibitem [{\citenamefont {Suhonen}(2007)}]{suhonen}%
  \BibitemOpen
  \bibfield  {author} {\bibinfo {author} {\bibfnamefont {J.}~\bibnamefont {Suhonen}},\ }\href {https://doi.org/10.1007/978-3-540-48861-3} {\emph {\bibinfo {title} {From Nucleons to Nucleus: Concepts of Microscopic Nuclear Theory}}}\ (\bibinfo  {publisher} {Springer Berlin},\ \bibinfo {address} {Heidelberg},\ \bibinfo {year} {2007})\BibitemShut {NoStop}%
\bibitem [{\citenamefont {Stetcu}\ \emph {et~al.}(2022{\natexlab{b}})\citenamefont {Stetcu}, \citenamefont {Baroni},\ and\ \citenamefont {Carlson}}]{6He}%
  \BibitemOpen
  \bibfield  {author} {\bibinfo {author} {\bibfnamefont {I.}~\bibnamefont {Stetcu}}, \bibinfo {author} {\bibfnamefont {A.}~\bibnamefont {Baroni}},\ and\ \bibinfo {author} {\bibfnamefont {J.}~\bibnamefont {Carlson}},\ }\href {https://doi.org/10.1103/PhysRevC.105.064308} {\bibfield  {journal} {\bibinfo  {journal} {Phys. Rev. C}\ }\textbf {\bibinfo {volume} {105}},\ \bibinfo {pages} {064308} (\bibinfo {year} {2022}{\natexlab{b}})}\BibitemShut {NoStop}%
\bibitem [{\citenamefont {Kiss}\ \emph {et~al.}(2022)\citenamefont {Kiss}, \citenamefont {Grossi}, \citenamefont {Lougovski}, \citenamefont {Sanchez}, \citenamefont {Vallecorsa},\ and\ \citenamefont {Papenbrock}}]{6Li}%
  \BibitemOpen
  \bibfield  {author} {\bibinfo {author} {\bibfnamefont {O.}~\bibnamefont {Kiss}}, \bibinfo {author} {\bibfnamefont {M.}~\bibnamefont {Grossi}}, \bibinfo {author} {\bibfnamefont {P.}~\bibnamefont {Lougovski}}, \bibinfo {author} {\bibfnamefont {F.}~\bibnamefont {Sanchez}}, \bibinfo {author} {\bibfnamefont {S.}~\bibnamefont {Vallecorsa}},\ and\ \bibinfo {author} {\bibfnamefont {T.}~\bibnamefont {Papenbrock}},\ }\href {https://doi.org/10.1103/PhysRevC.106.034325} {\bibfield  {journal} {\bibinfo  {journal} {Phys. Rev. C}\ }\textbf {\bibinfo {volume} {106}},\ \bibinfo {pages} {034325} (\bibinfo {year} {2022})}\BibitemShut {NoStop}%
\bibitem [{\citenamefont {P{\'e}rez-Obiol}\ \emph {et~al.}(2023)\citenamefont {P{\'e}rez-Obiol}, \citenamefont {Romero}, \citenamefont {Men{\'e}ndez}, \citenamefont {Rios}, \citenamefont {Garc{\'i}a-S{\'a}ez},\ and\ \citenamefont {Juli{\'a}-D{\'i}az}}]{perez-obiol_nuclear_2023}%
  \BibitemOpen
  \bibfield  {author} {\bibinfo {author} {\bibfnamefont {A.}~\bibnamefont {P{\'e}rez-Obiol}}, \bibinfo {author} {\bibfnamefont {A.~M.}\ \bibnamefont {Romero}}, \bibinfo {author} {\bibfnamefont {J.}~\bibnamefont {Men{\'e}ndez}}, \bibinfo {author} {\bibfnamefont {A.}~\bibnamefont {Rios}}, \bibinfo {author} {\bibfnamefont {A.}~\bibnamefont {Garc{\'i}a-S{\'a}ez}},\ and\ \bibinfo {author} {\bibfnamefont {B.}~\bibnamefont {Juli{\'a}-D{\'i}az}},\ }\href {https://doi.org/10.1038/s41598-023-39263-7} {\bibfield  {journal} {\bibinfo  {journal} {Scientific Reports}\ }\textbf {\bibinfo {volume} {13}},\ \bibinfo {pages} {12291} (\bibinfo {year} {2023})}\BibitemShut {NoStop}%
\bibitem [{\citenamefont {Sarma}\ \emph {et~al.}(2023)\citenamefont {Sarma}, \citenamefont {Di~Matteo}, \citenamefont {Abhishek},\ and\ \citenamefont {Srivastava}}]{sarma}%
  \BibitemOpen
  \bibfield  {author} {\bibinfo {author} {\bibfnamefont {C.}~\bibnamefont {Sarma}}, \bibinfo {author} {\bibfnamefont {O.}~\bibnamefont {Di~Matteo}}, \bibinfo {author} {\bibfnamefont {A.}~\bibnamefont {Abhishek}},\ and\ \bibinfo {author} {\bibfnamefont {P.~C.}\ \bibnamefont {Srivastava}},\ }\href {https://doi.org/10.1103/PhysRevC.108.064305} {\bibfield  {journal} {\bibinfo  {journal} {Phys. Rev. C}\ }\textbf {\bibinfo {volume} {108}},\ \bibinfo {pages} {064305} (\bibinfo {year} {2023})}\BibitemShut {NoStop}%
\bibitem [{\citenamefont {Massimi}\ \emph {et~al.}(2022)\citenamefont {Massimi}, \citenamefont {Cristallo}, \citenamefont {Domingo-Pardo},\ and\ \citenamefont {Lederer-Woods}}]{massimi_n_tof_2022}%
  \BibitemOpen
  \bibfield  {author} {\bibinfo {author} {\bibfnamefont {C.}~\bibnamefont {Massimi}}, \bibinfo {author} {\bibfnamefont {S.}~\bibnamefont {Cristallo}}, \bibinfo {author} {\bibfnamefont {C.}~\bibnamefont {Domingo-Pardo}},\ and\ \bibinfo {author} {\bibfnamefont {C.}~\bibnamefont {Lederer-Woods}},\ }\href {https://doi.org/10.3390/universe8020100} {\bibfield  {journal} {\bibinfo  {journal} {Universe}\ }\textbf {\bibinfo {volume} {8}},\ \bibinfo {pages} {100} (\bibinfo {year} {2022})}\BibitemShut {NoStop}%
\bibitem [{\citenamefont {{The n\_TOF Collaboration{\textdagger}}}\ \emph {et~al.}(2014)\citenamefont {{The n\_TOF Collaboration{\textdagger}}}, \citenamefont {{\v Z}ugec}, \citenamefont {Barbagallo}, \citenamefont {Colonna}, \citenamefont {Bosnar}, \citenamefont {Altstadt}, \citenamefont {Andrzejewski}, \citenamefont {Audouin}, \citenamefont {B{\'e}cares}, \citenamefont {Be{\v c}v{\'a}{\v r}}, \citenamefont {Belloni}, \citenamefont {Berthoumieux}, \citenamefont {Billowes}, \citenamefont {Boccone}, \citenamefont {Brugger}, \citenamefont {Calviani}, \citenamefont {Calvi{\~n}o}, \citenamefont {Cano-Ott}, \citenamefont {Carrapi{\c c}o}, \citenamefont {Cerutti}, \citenamefont {Chiaveri}, \citenamefont {Chin}, \citenamefont {Cort{\'e}s}, \citenamefont {Cort{\'e}s-Giraldo}, \citenamefont {Diakaki}, \citenamefont {Domingo-Pardo}, \citenamefont {Duran}, \citenamefont {Dzysiuk}, \citenamefont {Eleftheriadis}, \citenamefont {Ferrari}, \citenamefont {Fraval}, \citenamefont {Ganesan}, \citenamefont {Garc{\'i}a},
  \citenamefont {Giubrone}, \citenamefont {G{\'o}mez-Hornillos}, \citenamefont {Gon{\c c}alves}, \citenamefont {Gonz{\'a}lez-Romero}, \citenamefont {Griesmayer}, \citenamefont {Guerrero}, \citenamefont {Gunsing}, \citenamefont {Gurusamy}, \citenamefont {Jenkins}, \citenamefont {Jericha}, \citenamefont {Kadi}, \citenamefont {K{\"a}ppeler}, \citenamefont {Karadimos}, \citenamefont {Koehler}, \citenamefont {Kokkoris}, \citenamefont {Krti{\v c}ka}, \citenamefont {Kroll}, \citenamefont {Langer}, \citenamefont {Lederer}, \citenamefont {Leeb}, \citenamefont {Leong}, \citenamefont {Losito}, \citenamefont {Manousos}, \citenamefont {Marganiec}, \citenamefont {Mart{\'i}nez}, \citenamefont {Massimi}, \citenamefont {Mastinu}, \citenamefont {Mastromarco}, \citenamefont {Meaze}, \citenamefont {Mendoza}, \citenamefont {Mengoni}, \citenamefont {Milazzo}, \citenamefont {Mingrone}, \citenamefont {Mirea}, \citenamefont {Mondalaers}, \citenamefont {Paradela}, \citenamefont {Pavlik}, \citenamefont {Perkowski}, \citenamefont
  {Pignatari}, \citenamefont {Plompen}, \citenamefont {Praena}, \citenamefont {Quesada}, \citenamefont {Rauscher}, \citenamefont {Reifarth}, \citenamefont {Riego}, \citenamefont {Roman}, \citenamefont {Rubbia}, \citenamefont {Sarmento}, \citenamefont {Schillebeeckx}, \citenamefont {Schmidt}, \citenamefont {Tagliente}, \citenamefont {Tain}, \citenamefont {Tarr{\'i}o}, \citenamefont {Tassan-Got}, \citenamefont {Tsinganis}, \citenamefont {Valenta}, \citenamefont {Vannini}, \citenamefont {Variale}, \citenamefont {Vaz}, \citenamefont {Ventura}, \citenamefont {Versaci}, \citenamefont {Vermeulen}, \citenamefont {Vlachoudis}, \citenamefont {Vlastou}, \citenamefont {Wallner}, \citenamefont {Ware}, \citenamefont {Weigand}, \citenamefont {Wei{\ss}},\ and\ \citenamefont {Wright}}]{the_n_tof_collaboration_experimental_2014}%
  \BibitemOpen
  \bibfield  {author} {\bibinfo {author} {\bibnamefont {{The n\_TOF Collaboration{\textdagger}}}}, \bibinfo {author} {\bibfnamefont {P.}~\bibnamefont {{\v Z}ugec}}, \bibinfo {author} {\bibfnamefont {M.}~\bibnamefont {Barbagallo}}, \bibinfo {author} {\bibfnamefont {N.}~\bibnamefont {Colonna}}, \bibinfo {author} {\bibfnamefont {D.}~\bibnamefont {Bosnar}}, \bibinfo {author} {\bibfnamefont {S.}~\bibnamefont {Altstadt}}, \bibinfo {author} {\bibfnamefont {J.}~\bibnamefont {Andrzejewski}}, \bibinfo {author} {\bibfnamefont {L.}~\bibnamefont {Audouin}}, \bibinfo {author} {\bibfnamefont {V.}~\bibnamefont {B{\'e}cares}}, \bibinfo {author} {\bibfnamefont {F.}~\bibnamefont {Be{\v c}v{\'a}{\v r}}}, \bibinfo {author} {\bibfnamefont {F.}~\bibnamefont {Belloni}}, \bibinfo {author} {\bibfnamefont {E.}~\bibnamefont {Berthoumieux}}, \bibinfo {author} {\bibfnamefont {J.}~\bibnamefont {Billowes}}, \bibinfo {author} {\bibfnamefont {V.}~\bibnamefont {Boccone}}, \bibinfo {author} {\bibfnamefont {M.}~\bibnamefont {Brugger}}, \bibinfo
  {author} {\bibfnamefont {M.}~\bibnamefont {Calviani}}, \bibinfo {author} {\bibfnamefont {F.}~\bibnamefont {Calvi{\~n}o}}, \bibinfo {author} {\bibfnamefont {D.}~\bibnamefont {Cano-Ott}}, \bibinfo {author} {\bibfnamefont {C.}~\bibnamefont {Carrapi{\c c}o}}, \bibinfo {author} {\bibfnamefont {F.}~\bibnamefont {Cerutti}}, \bibinfo {author} {\bibfnamefont {E.}~\bibnamefont {Chiaveri}}, \bibinfo {author} {\bibfnamefont {M.}~\bibnamefont {Chin}}, \bibinfo {author} {\bibfnamefont {G.}~\bibnamefont {Cort{\'e}s}}, \bibinfo {author} {\bibfnamefont {M.~A.}\ \bibnamefont {Cort{\'e}s-Giraldo}}, \bibinfo {author} {\bibfnamefont {M.}~\bibnamefont {Diakaki}}, \bibinfo {author} {\bibfnamefont {C.}~\bibnamefont {Domingo-Pardo}}, \bibinfo {author} {\bibfnamefont {I.}~\bibnamefont {Duran}}, \bibinfo {author} {\bibfnamefont {N.}~\bibnamefont {Dzysiuk}}, \bibinfo {author} {\bibfnamefont {C.}~\bibnamefont {Eleftheriadis}}, \bibinfo {author} {\bibfnamefont {A.}~\bibnamefont {Ferrari}}, \bibinfo {author} {\bibfnamefont
  {K.}~\bibnamefont {Fraval}}, \bibinfo {author} {\bibfnamefont {S.}~\bibnamefont {Ganesan}}, \bibinfo {author} {\bibfnamefont {A.~R.}\ \bibnamefont {Garc{\'i}a}}, \bibinfo {author} {\bibfnamefont {G.}~\bibnamefont {Giubrone}}, \bibinfo {author} {\bibfnamefont {M.~B.}\ \bibnamefont {G{\'o}mez-Hornillos}}, \bibinfo {author} {\bibfnamefont {I.~F.}\ \bibnamefont {Gon{\c c}alves}}, \bibinfo {author} {\bibfnamefont {E.}~\bibnamefont {Gonz{\'a}lez-Romero}}, \bibinfo {author} {\bibfnamefont {E.}~\bibnamefont {Griesmayer}}, \bibinfo {author} {\bibfnamefont {C.}~\bibnamefont {Guerrero}}, \bibinfo {author} {\bibfnamefont {F.}~\bibnamefont {Gunsing}}, \bibinfo {author} {\bibfnamefont {P.}~\bibnamefont {Gurusamy}}, \bibinfo {author} {\bibfnamefont {D.~G.}\ \bibnamefont {Jenkins}}, \bibinfo {author} {\bibfnamefont {E.}~\bibnamefont {Jericha}}, \bibinfo {author} {\bibfnamefont {Y.}~\bibnamefont {Kadi}}, \bibinfo {author} {\bibfnamefont {F.}~\bibnamefont {K{\"a}ppeler}}, \bibinfo {author} {\bibfnamefont {D.}~\bibnamefont
  {Karadimos}}, \bibinfo {author} {\bibfnamefont {P.}~\bibnamefont {Koehler}}, \bibinfo {author} {\bibfnamefont {M.}~\bibnamefont {Kokkoris}}, \bibinfo {author} {\bibfnamefont {M.}~\bibnamefont {Krti{\v c}ka}}, \bibinfo {author} {\bibfnamefont {J.}~\bibnamefont {Kroll}}, \bibinfo {author} {\bibfnamefont {C.}~\bibnamefont {Langer}}, \bibinfo {author} {\bibfnamefont {C.}~\bibnamefont {Lederer}}, \bibinfo {author} {\bibfnamefont {H.}~\bibnamefont {Leeb}}, \bibinfo {author} {\bibfnamefont {L.~S.}\ \bibnamefont {Leong}}, \bibinfo {author} {\bibfnamefont {R.}~\bibnamefont {Losito}}, \bibinfo {author} {\bibfnamefont {A.}~\bibnamefont {Manousos}}, \bibinfo {author} {\bibfnamefont {J.}~\bibnamefont {Marganiec}}, \bibinfo {author} {\bibfnamefont {T.}~\bibnamefont {Mart{\'i}nez}}, \bibinfo {author} {\bibfnamefont {C.}~\bibnamefont {Massimi}}, \bibinfo {author} {\bibfnamefont {P.~F.}\ \bibnamefont {Mastinu}}, \bibinfo {author} {\bibfnamefont {M.}~\bibnamefont {Mastromarco}}, \bibinfo {author} {\bibfnamefont
  {M.}~\bibnamefont {Meaze}}, \bibinfo {author} {\bibfnamefont {E.}~\bibnamefont {Mendoza}}, \bibinfo {author} {\bibfnamefont {A.}~\bibnamefont {Mengoni}}, \bibinfo {author} {\bibfnamefont {P.~M.}\ \bibnamefont {Milazzo}}, \bibinfo {author} {\bibfnamefont {F.}~\bibnamefont {Mingrone}}, \bibinfo {author} {\bibfnamefont {M.}~\bibnamefont {Mirea}}, \bibinfo {author} {\bibfnamefont {W.}~\bibnamefont {Mondalaers}}, \bibinfo {author} {\bibfnamefont {C.}~\bibnamefont {Paradela}}, \bibinfo {author} {\bibfnamefont {A.}~\bibnamefont {Pavlik}}, \bibinfo {author} {\bibfnamefont {J.}~\bibnamefont {Perkowski}}, \bibinfo {author} {\bibfnamefont {M.}~\bibnamefont {Pignatari}}, \bibinfo {author} {\bibfnamefont {A.}~\bibnamefont {Plompen}}, \bibinfo {author} {\bibfnamefont {J.}~\bibnamefont {Praena}}, \bibinfo {author} {\bibfnamefont {J.~M.}\ \bibnamefont {Quesada}}, \bibinfo {author} {\bibfnamefont {T.}~\bibnamefont {Rauscher}}, \bibinfo {author} {\bibfnamefont {R.}~\bibnamefont {Reifarth}}, \bibinfo {author} {\bibfnamefont
  {A.}~\bibnamefont {Riego}}, \bibinfo {author} {\bibfnamefont {F.}~\bibnamefont {Roman}}, \bibinfo {author} {\bibfnamefont {C.}~\bibnamefont {Rubbia}}, \bibinfo {author} {\bibfnamefont {R.}~\bibnamefont {Sarmento}}, \bibinfo {author} {\bibfnamefont {P.}~\bibnamefont {Schillebeeckx}}, \bibinfo {author} {\bibfnamefont {S.}~\bibnamefont {Schmidt}}, \bibinfo {author} {\bibfnamefont {G.}~\bibnamefont {Tagliente}}, \bibinfo {author} {\bibfnamefont {J.~L.}\ \bibnamefont {Tain}}, \bibinfo {author} {\bibfnamefont {D.}~\bibnamefont {Tarr{\'i}o}}, \bibinfo {author} {\bibfnamefont {L.}~\bibnamefont {Tassan-Got}}, \bibinfo {author} {\bibfnamefont {A.}~\bibnamefont {Tsinganis}}, \bibinfo {author} {\bibfnamefont {S.}~\bibnamefont {Valenta}}, \bibinfo {author} {\bibfnamefont {G.}~\bibnamefont {Vannini}}, \bibinfo {author} {\bibfnamefont {V.}~\bibnamefont {Variale}}, \bibinfo {author} {\bibfnamefont {P.}~\bibnamefont {Vaz}}, \bibinfo {author} {\bibfnamefont {A.}~\bibnamefont {Ventura}}, \bibinfo {author} {\bibfnamefont
  {R.}~\bibnamefont {Versaci}}, \bibinfo {author} {\bibfnamefont {M.~J.}\ \bibnamefont {Vermeulen}}, \bibinfo {author} {\bibfnamefont {V.}~\bibnamefont {Vlachoudis}}, \bibinfo {author} {\bibfnamefont {R.}~\bibnamefont {Vlastou}}, \bibinfo {author} {\bibfnamefont {A.}~\bibnamefont {Wallner}}, \bibinfo {author} {\bibfnamefont {T.}~\bibnamefont {Ware}}, \bibinfo {author} {\bibfnamefont {M.}~\bibnamefont {Weigand}}, \bibinfo {author} {\bibfnamefont {C.}~\bibnamefont {Wei{\ss}}},\ and\ \bibinfo {author} {\bibfnamefont {T.}~\bibnamefont {Wright}},\ }\href {https://doi.org/10.1103/PhysRevC.89.014605} {\bibfield  {journal} {\bibinfo  {journal} {Physical Review C}\ }\textbf {\bibinfo {volume} {89}},\ \bibinfo {pages} {014605} (\bibinfo {year} {2014})}\BibitemShut {NoStop}%
\bibitem [{\citenamefont {Guber}\ \emph {et~al.}(2010)\citenamefont {Guber}, \citenamefont {Derrien}, \citenamefont {Leal}, \citenamefont {Arbanas}, \citenamefont {Wiarda}, \citenamefont {Koehler},\ and\ \citenamefont {Harvey}}]{guber_astrophysical_2010}%
  \BibitemOpen
  \bibfield  {author} {\bibinfo {author} {\bibfnamefont {K.~H.}\ \bibnamefont {Guber}}, \bibinfo {author} {\bibfnamefont {H.}~\bibnamefont {Derrien}}, \bibinfo {author} {\bibfnamefont {L.~C.}\ \bibnamefont {Leal}}, \bibinfo {author} {\bibfnamefont {G.}~\bibnamefont {Arbanas}}, \bibinfo {author} {\bibfnamefont {D.}~\bibnamefont {Wiarda}}, \bibinfo {author} {\bibfnamefont {P.~E.}\ \bibnamefont {Koehler}},\ and\ \bibinfo {author} {\bibfnamefont {J.~A.}\ \bibnamefont {Harvey}},\ }\href {https://doi.org/10.1103/PhysRevC.82.057601} {\bibfield  {journal} {\bibinfo  {journal} {Physical Review C}\ }\textbf {\bibinfo {volume} {82}},\ \bibinfo {pages} {057601} (\bibinfo {year} {2010})}\BibitemShut {NoStop}%
\bibitem [{\citenamefont {Alhassan}\ \emph {et~al.}(2024)\citenamefont {Alhassan}, \citenamefont {Rochman}, \citenamefont {Schnabel},\ and\ \citenamefont {Koning}}]{alhassan_bayesian_2024}%
  \BibitemOpen
  \bibfield  {author} {\bibinfo {author} {\bibfnamefont {E.}~\bibnamefont {Alhassan}}, \bibinfo {author} {\bibfnamefont {D.}~\bibnamefont {Rochman}}, \bibinfo {author} {\bibfnamefont {G.}~\bibnamefont {Schnabel}},\ and\ \bibinfo {author} {\bibfnamefont {A.~J.}\ \bibnamefont {Koning}},\ }\href {https://doi.org/10.48550/arXiv.2402.14363} {\bibinfo {title} {Bayesian {Model} {Averaging} ({BMA}) for nuclear data evaluation}} (\bibinfo {year} {2024}),\ \bibinfo {note} {arXiv:2402.14363 [nucl-th]}\BibitemShut {NoStop}%
\bibitem [{\citenamefont {Luneville}\ \emph {et~al.}(2018)\citenamefont {Luneville}, \citenamefont {Sublet},\ and\ \citenamefont {Simeone}}]{luneville_impact_2018}%
  \BibitemOpen
  \bibfield  {author} {\bibinfo {author} {\bibfnamefont {L.}~\bibnamefont {Luneville}}, \bibinfo {author} {\bibfnamefont {J.~C.}\ \bibnamefont {Sublet}},\ and\ \bibinfo {author} {\bibfnamefont {D.}~\bibnamefont {Simeone}},\ }\href {https://doi.org/10.1016/j.jnucmat.2017.06.039} {\bibfield  {journal} {\bibinfo  {journal} {Journal of Nuclear Materials}\ }\textbf {\bibinfo {volume} {505}},\ \bibinfo {pages} {262} (\bibinfo {year} {2018})}\BibitemShut {NoStop}%
\bibitem [{\citenamefont {Honma}\ \emph {et~al.}(2009)\citenamefont {Honma}, \citenamefont {Otsuka}, \citenamefont {Mizusaki},\ and\ \citenamefont {Hjorth-Jensen}}]{jun45}%
  \BibitemOpen
  \bibfield  {author} {\bibinfo {author} {\bibfnamefont {M.}~\bibnamefont {Honma}}, \bibinfo {author} {\bibfnamefont {T.}~\bibnamefont {Otsuka}}, \bibinfo {author} {\bibfnamefont {T.}~\bibnamefont {Mizusaki}},\ and\ \bibinfo {author} {\bibfnamefont {M.}~\bibnamefont {Hjorth-Jensen}},\ }\href {https://doi.org/10.1103/PhysRevC.80.064323} {\bibfield  {journal} {\bibinfo  {journal} {Phys. Rev. C}\ }\textbf {\bibinfo {volume} {80}},\ \bibinfo {pages} {064323} (\bibinfo {year} {2009})}\BibitemShut {NoStop}%
\bibitem [{\citenamefont {Jordan}\ and\ \citenamefont {Wigner}(1928)}]{JW}%
  \BibitemOpen
  \bibfield  {author} {\bibinfo {author} {\bibfnamefont {P.}~\bibnamefont {Jordan}}\ and\ \bibinfo {author} {\bibfnamefont {E.}~\bibnamefont {Wigner}},\ }\href {https://doi.org/10.1007/BF01331938} {\bibfield  {journal} {\bibinfo  {journal} {Zeitschrift für Physik}\ }\textbf {\bibinfo {volume} {47}},\ \bibinfo {pages} {631} (\bibinfo {year} {1928})}\BibitemShut {NoStop}%
\bibitem [{\citenamefont {Siwach}\ and\ \citenamefont {Arumugam}(2021{\natexlab{b}})}]{PhysRevC.104.034301}%
  \BibitemOpen
  \bibfield  {author} {\bibinfo {author} {\bibfnamefont {P.}~\bibnamefont {Siwach}}\ and\ \bibinfo {author} {\bibfnamefont {P.}~\bibnamefont {Arumugam}},\ }\href {https://doi.org/10.1103/PhysRevC.104.034301} {\bibfield  {journal} {\bibinfo  {journal} {Phys. Rev. C}\ }\textbf {\bibinfo {volume} {104}},\ \bibinfo {pages} {034301} (\bibinfo {year} {2021}{\natexlab{b}})}\BibitemShut {NoStop}%
\bibitem [{\citenamefont {Bravyi}\ and\ \citenamefont {Kitaev}(2002)}]{BK}%
  \BibitemOpen
  \bibfield  {author} {\bibinfo {author} {\bibfnamefont {S.~B.}\ \bibnamefont {Bravyi}}\ and\ \bibinfo {author} {\bibfnamefont {A.~Y.}\ \bibnamefont {Kitaev}},\ }\href {https://doi.org/https://doi.org/10.1006/aphy.2002.6254} {\bibfield  {journal} {\bibinfo  {journal} {Annals of Physics}\ }\textbf {\bibinfo {volume} {298}},\ \bibinfo {pages} {210} (\bibinfo {year} {2002})}\BibitemShut {NoStop}%
\bibitem [{\citenamefont {Peruzzo}\ \emph {et~al.}(2014)\citenamefont {Peruzzo}, \citenamefont {McClean}, \citenamefont {Shadbolt}, \citenamefont {Yung}, \citenamefont {Zhou}, \citenamefont {Love}, \citenamefont {Aspuru-Guzik},\ and\ \citenamefont {O’Brien}}]{peruzzo}%
  \BibitemOpen
  \bibfield  {author} {\bibinfo {author} {\bibfnamefont {A.}~\bibnamefont {Peruzzo}}, \bibinfo {author} {\bibfnamefont {J.}~\bibnamefont {McClean}}, \bibinfo {author} {\bibfnamefont {P.}~\bibnamefont {Shadbolt}}, \bibinfo {author} {\bibfnamefont {M.-H.}\ \bibnamefont {Yung}}, \bibinfo {author} {\bibfnamefont {X.-Q.}\ \bibnamefont {Zhou}}, \bibinfo {author} {\bibfnamefont {P.~J.}\ \bibnamefont {Love}}, \bibinfo {author} {\bibfnamefont {A.}~\bibnamefont {Aspuru-Guzik}},\ and\ \bibinfo {author} {\bibfnamefont {J.~L.}\ \bibnamefont {O’Brien}},\ }\href {https://doi.org/10.1038/ncomms5213} {\bibfield  {journal} {\bibinfo  {journal} {Nature Communications}\ }\textbf {\bibinfo {volume} {5}},\ \bibinfo {pages} {4213} (\bibinfo {year} {2014})}\BibitemShut {NoStop}%
\bibitem [{\citenamefont {Cerezo}\ \emph {et~al.}(2021)\citenamefont {Cerezo}, \citenamefont {Arrasmith}, \citenamefont {Babbush}, \citenamefont {Benjamin}, \citenamefont {Endo}, \citenamefont {Fujii}, \citenamefont {McClean}, \citenamefont {Mitarai}, \citenamefont {Yuan}, \citenamefont {Cincio},\ and\ \citenamefont {Coles}}]{cerezo2020variationalreview}%
  \BibitemOpen
  \bibfield  {author} {\bibinfo {author} {\bibfnamefont {M.}~\bibnamefont {Cerezo}}, \bibinfo {author} {\bibfnamefont {A.}~\bibnamefont {Arrasmith}}, \bibinfo {author} {\bibfnamefont {R.}~\bibnamefont {Babbush}}, \bibinfo {author} {\bibfnamefont {S.~C.}\ \bibnamefont {Benjamin}}, \bibinfo {author} {\bibfnamefont {S.}~\bibnamefont {Endo}}, \bibinfo {author} {\bibfnamefont {K.}~\bibnamefont {Fujii}}, \bibinfo {author} {\bibfnamefont {J.~R.}\ \bibnamefont {McClean}}, \bibinfo {author} {\bibfnamefont {K.}~\bibnamefont {Mitarai}}, \bibinfo {author} {\bibfnamefont {X.}~\bibnamefont {Yuan}}, \bibinfo {author} {\bibfnamefont {L.}~\bibnamefont {Cincio}},\ and\ \bibinfo {author} {\bibfnamefont {P.~J.}\ \bibnamefont {Coles}},\ }\href {https://doi.org/10.1038/s42254-021-00348-9} {\bibfield  {journal} {\bibinfo  {journal} {Nature Reviews Physics}\ }\textbf {\bibinfo {volume} {3}},\ \bibinfo {pages} {625–644} (\bibinfo {year} {2021})}\BibitemShut {NoStop}%
\bibitem [{\citenamefont {Fedorov}\ \emph {et~al.}(2022)\citenamefont {Fedorov}, \citenamefont {Peng}, \citenamefont {Govind},\ and\ \citenamefont {Alexeev}}]{fedorov_vqe_2022}%
  \BibitemOpen
  \bibfield  {author} {\bibinfo {author} {\bibfnamefont {D.~A.}\ \bibnamefont {Fedorov}}, \bibinfo {author} {\bibfnamefont {B.}~\bibnamefont {Peng}}, \bibinfo {author} {\bibfnamefont {N.}~\bibnamefont {Govind}},\ and\ \bibinfo {author} {\bibfnamefont {Y.}~\bibnamefont {Alexeev}},\ }\href {https://doi.org/10.1186/s41313-021-00032-6} {\bibfield  {journal} {\bibinfo  {journal} {Materials Theory}\ }\textbf {\bibinfo {volume} {6}},\ \bibinfo {pages} {2} (\bibinfo {year} {2022})}\BibitemShut {NoStop}%
\bibitem [{\citenamefont {Hobday}\ \emph {et~al.}(2023)\citenamefont {Hobday}, \citenamefont {Stevenson},\ and\ \citenamefont {Benstead}}]{hobday_variance_2023}%
  \BibitemOpen
  \bibfield  {author} {\bibinfo {author} {\bibfnamefont {I.}~\bibnamefont {Hobday}}, \bibinfo {author} {\bibfnamefont {P.}~\bibnamefont {Stevenson}},\ and\ \bibinfo {author} {\bibfnamefont {J.}~\bibnamefont {Benstead}},\ }\href {https://doi.org/10.1051/epjconf/202328416002} {\bibfield  {journal} {\bibinfo  {journal} {EPJ Web of Conferences}\ }\textbf {\bibinfo {volume} {284}},\ \bibinfo {pages} {16002} (\bibinfo {year} {2023})},\ \bibinfo {note} {publisher: EDP Sciences}\BibitemShut {NoStop}%
\bibitem [{\citenamefont {Coester}(1958)}]{coester_bound_1958}%
  \BibitemOpen
  \bibfield  {author} {\bibinfo {author} {\bibfnamefont {F.}~\bibnamefont {Coester}},\ }\href {https://doi.org/10.1016/0029-5582(58)90280-3} {\bibfield  {journal} {\bibinfo  {journal} {Nuclear Physics}\ }\textbf {\bibinfo {volume} {7}},\ \bibinfo {pages} {421} (\bibinfo {year} {1958})}\BibitemShut {NoStop}%
\bibitem [{\citenamefont {K{\"u}mmel}\ \emph {et~al.}(1978)\citenamefont {K{\"u}mmel}, \citenamefont {L{\"u}hrmann},\ and\ \citenamefont {Zabolitzky}}]{kummel_many-fermion_1978}%
  \BibitemOpen
  \bibfield  {author} {\bibinfo {author} {\bibfnamefont {H.}~\bibnamefont {K{\"u}mmel}}, \bibinfo {author} {\bibfnamefont {K.~H.}\ \bibnamefont {L{\"u}hrmann}},\ and\ \bibinfo {author} {\bibfnamefont {J.~G.}\ \bibnamefont {Zabolitzky}},\ }\href {https://doi.org/10.1016/0370-1573(78)90081-9} {\bibfield  {journal} {\bibinfo  {journal} {Physics Reports}\ }\textbf {\bibinfo {volume} {36}},\ \bibinfo {pages} {1} (\bibinfo {year} {1978})}\BibitemShut {NoStop}%
\bibitem [{\citenamefont {Arrazola}\ \emph {et~al.}(2022)\citenamefont {Arrazola}, \citenamefont {Di~Matteo}, \citenamefont {Quesada}, \citenamefont {Jahangiri}, \citenamefont {Delgado},\ and\ \citenamefont {Killoran}}]{givens_rot}%
  \BibitemOpen
  \bibfield  {author} {\bibinfo {author} {\bibfnamefont {J.~M.}\ \bibnamefont {Arrazola}}, \bibinfo {author} {\bibfnamefont {O.}~\bibnamefont {Di~Matteo}}, \bibinfo {author} {\bibfnamefont {N.}~\bibnamefont {Quesada}}, \bibinfo {author} {\bibfnamefont {S.}~\bibnamefont {Jahangiri}}, \bibinfo {author} {\bibfnamefont {A.}~\bibnamefont {Delgado}},\ and\ \bibinfo {author} {\bibfnamefont {N.}~\bibnamefont {Killoran}},\ }\href {https://doi.org/10.22331/q-2022-06-20-742} {\bibfield  {journal} {\bibinfo  {journal} {{Quantum}}\ }\textbf {\bibinfo {volume} {6}},\ \bibinfo {pages} {742} (\bibinfo {year} {2022})}\BibitemShut {NoStop}%
\bibitem [{\citenamefont {Anselmetti}\ \emph {et~al.}(2021)\citenamefont {Anselmetti}, \citenamefont {Wierichs}, \citenamefont {Gogolin},\ and\ \citenamefont {Parrish}}]{G2}%
  \BibitemOpen
  \bibfield  {author} {\bibinfo {author} {\bibfnamefont {G.-L.~R.}\ \bibnamefont {Anselmetti}}, \bibinfo {author} {\bibfnamefont {D.}~\bibnamefont {Wierichs}}, \bibinfo {author} {\bibfnamefont {C.}~\bibnamefont {Gogolin}},\ and\ \bibinfo {author} {\bibfnamefont {R.~M.}\ \bibnamefont {Parrish}},\ }\href {https://doi.org/10.1088/1367-2630/ac2cb3} {\bibfield  {journal} {\bibinfo  {journal} {New Journal of Physics}\ }\textbf {\bibinfo {volume} {23}},\ \bibinfo {pages} {113010} (\bibinfo {year} {2021})}\BibitemShut {NoStop}%
\bibitem [{\citenamefont {{Qiskit contributors}}(2023)}]{Qiskit}%
  \BibitemOpen
  \bibfield  {author} {\bibinfo {author} {\bibnamefont {{Qiskit contributors}}},\ }\href {https://doi.org/10.5281/zenodo.2573505} {\bibinfo {title} {Qiskit: An open-source framework for quantum computing}} (\bibinfo {year} {2023})\BibitemShut {NoStop}%
\bibitem [{\citenamefont {Powell}(1994)}]{cobyla}%
  \BibitemOpen
  \bibfield  {author} {\bibinfo {author} {\bibfnamefont {M.~J.~D.}\ \bibnamefont {Powell}},\ }\bibinfo {title} {A direct search optimization method that models the objective and constraint functions by linear interpolation},\ in\ \href {https://doi.org/10.1007/978-94-015-8330-5_4} {\emph {\bibinfo {booktitle} {Advances in Optimization and Numerical Analysis}}}\ (\bibinfo  {publisher} {Springer Netherlands},\ \bibinfo {address} {Dordrecht},\ \bibinfo {year} {1994})\ pp.\ \bibinfo {pages} {51--67}\BibitemShut {NoStop}%
\bibitem [{\citenamefont {Kraft}(1988)}]{slsqp}%
  \BibitemOpen
  \bibfield  {author} {\bibinfo {author} {\bibfnamefont {D.}~\bibnamefont {Kraft}},\ }\href {https://books.google.co.uk/books?id=4rKaGwAACAAJ} {\emph {\bibinfo {title} {A Software Package for Sequential Quadratic Programming}}},\ Deutsche Forschungs- und Versuchsanstalt f{\"u}r Luft- und Raumfahrt K{\"o}ln: Forschungsbericht\ (\bibinfo  {publisher} {Wiss. Berichtswesen d. DFVLR},\ \bibinfo {year} {1988})\BibitemShut {NoStop}%
\bibitem [{\citenamefont {Spall}(1992)}]{spsa}%
  \BibitemOpen
  \bibfield  {author} {\bibinfo {author} {\bibfnamefont {J.}~\bibnamefont {Spall}},\ }\href {https://doi.org/10.1109/9.119632} {\bibfield  {journal} {\bibinfo  {journal} {IEEE Transactions on Automatic Control}\ }\textbf {\bibinfo {volume} {37}},\ \bibinfo {pages} {332} (\bibinfo {year} {1992})}\BibitemShut {NoStop}%
\bibitem [{\citenamefont {Holmes}\ \emph {et~al.}(2022)\citenamefont {Holmes}, \citenamefont {Sharma}, \citenamefont {Cerezo},\ and\ \citenamefont {Coles}}]{PRXQuantum.3.010313}%
  \BibitemOpen
  \bibfield  {author} {\bibinfo {author} {\bibfnamefont {Z.}~\bibnamefont {Holmes}}, \bibinfo {author} {\bibfnamefont {K.}~\bibnamefont {Sharma}}, \bibinfo {author} {\bibfnamefont {M.}~\bibnamefont {Cerezo}},\ and\ \bibinfo {author} {\bibfnamefont {P.~J.}\ \bibnamefont {Coles}},\ }\href {https://doi.org/10.1103/PRXQuantum.3.010313} {\bibfield  {journal} {\bibinfo  {journal} {PRX Quantum}\ }\textbf {\bibinfo {volume} {3}},\ \bibinfo {pages} {010313} (\bibinfo {year} {2022})}\BibitemShut {NoStop}%
\bibitem [{\citenamefont {Kandala}\ \emph {et~al.}(2017)\citenamefont {Kandala}, \citenamefont {Mezzacapo}, \citenamefont {Temme}, \citenamefont {Takita}, \citenamefont {Brink}, \citenamefont {Chow},\ and\ \citenamefont {Gambetta}}]{spsa2}%
  \BibitemOpen
  \bibfield  {author} {\bibinfo {author} {\bibfnamefont {A.}~\bibnamefont {Kandala}}, \bibinfo {author} {\bibfnamefont {A.}~\bibnamefont {Mezzacapo}}, \bibinfo {author} {\bibfnamefont {K.}~\bibnamefont {Temme}}, \bibinfo {author} {\bibfnamefont {M.}~\bibnamefont {Takita}}, \bibinfo {author} {\bibfnamefont {M.}~\bibnamefont {Brink}}, \bibinfo {author} {\bibfnamefont {J.~M.}\ \bibnamefont {Chow}},\ and\ \bibinfo {author} {\bibfnamefont {J.~M.}\ \bibnamefont {Gambetta}},\ }\href {https://doi.org/10.1038/nature23879} {\bibfield  {journal} {\bibinfo  {journal} {Nature}\ }\textbf {\bibinfo {volume} {549}},\ \bibinfo {pages} {242} (\bibinfo {year} {2017})}\BibitemShut {NoStop}%
\end{thebibliography}%

\end{document}